\documentclass[aps,showpacs,twocolumn,showkeys,tightenlines,floatfix,superscriptaddress]{revtex4}
\usepackage{epsfig,amssymb,bm,graphics,color}
\newcommand*{\fs}[1]{#1\!\!\!/}

\newcommand*{\ee}{e^+e^-}

\begin{document} {\normalsize }
\title{Determination of the carrier envelope phase for short, circularly polarized laser pulses}
\author{
 Alexander~I.~Titov}
  \affiliation{
 Bogoliubov Laboratory of Theoretical Physics, JINR, Dubna 141980, Russia}
 \author{Burkhard~K\"ampfer}
 \affiliation{Helmholtz-Zentrum  Dresden-Rossendorf, 01314 Dresden, Germany}
 \affiliation{Institut f\"ur Theoretische Physik, TU~Dresden, 01062 Dresden, Germany}
 \author{ Atsushi Hosaka}
 \affiliation{RCNP, 10-1 Mihogaoka Ibaraki, 567-0047 Osaka, Japan}
 \affiliation{J-PARC Branch, KEK, Tokai, Ibaraki, 319-1106, Japan}
 \author{Tobias~Nousch}
 \affiliation{Institut f\"ur Theoretische Physik, TU~Dresden, 01062 Dresden, Germany}
 \author{Daniel~Seipt}
 \affiliation{Helmholtz-Institut Jena, Fr\"obelstieg 3, 07743 Jena, Germany}
\begin{abstract}
We analyze the impact of the carrier envelope phase on the
differential cross sections of the Breit-Wheeler and the
generalized Compton scattering in the interaction of a charged electron
(positron) with an intensive ultra-short electromagnetic (laser)
pulse. The differential cross sections as a function of the azimuthal
angle of the outgoing electron have a clear bump structure, where the bump
position coincides with the value of the carrier phase. This effect can
be used for the carrier envelope phase determination.
\end{abstract}
\pacs{12.20.Ds, 13.40.-f, 23.20.Nx} \keywords{Carrier envelope
phase, non-linear QED dynamics, multi-photon, sub-threshold
processes} \maketitle
\section{introduction}

 {The rapid progress in laser technology \cite{Tajima}
 offers novel and unprecedented opportunities to investigate
 quantum systems with intense laser beams~\cite{Piazza}.}
 A laser intensity $I_L$ of $\sim 2\times 10^{22}$  W/cm${}^2$ has been already
 achieved~\cite{I-22}. Intensities of the order of
 $I_L \sim 10^{23}...10^{25}$ W/cm$^2$ are envisaged in the near future, e.g.\
 at  CLF~\cite{CLF}, ELI~\cite{ELI}, or HiPER~\cite{hiper}.
 Further facilities are in the planning or construction stages, e.g.
 the PEARL laser facility~\cite{sarov} at Sarov/Nizhny Novgorod, Russia.
 The high intensities are provided in short
 pulses on a femtosecond pulse duration
 level~\cite{Piazza,ShortPulse,ShortPulse_2},
 with only a few oscillations of the electromagnetic (e.m.) field
 or even sub-cycle pulses.
 (The tight connection of high intensity and short pulse duration
 is further emphasized in \cite{Mackenroth-2011}. The attosecond
 regime will become accessible at shorter wavelengths~\cite{atto,I-222}).

 The new laser facilities utilize
 short and ultra-short pulses in "one-" or "few-" cycle regimes.
 In this case, a determination of the pulse fine-structure is very important
 and, in particular, tasking the phase difference between the electric
 field and pulse envelope, i.~e. the carrier envelope phase (CEP).
 In the past, in the case of low beam intensity,
 the CEP determination has been achieved by
 averaging over a large number of phase-stabilized laser pulses.
 Later on, some methods for a determination of CEP,
 were elaborated.
 For example,
 by studying above-threshold ion-ionization~\cite{CE1,CE2},
 a direct measurement of the light waves of visible, ultraviolet,
 and/or infrared light using an electron atto-second probe~\cite{CE3}.
 {As another possibility for CEP measurement it has been established to
 convert the light to the terahertz (THz)
 frequency range in a plasma, with subsequent analysis of the spatial charge asymmetry
 associated with THz emission~\cite{CE4}.}

 However, such methods can not
 be applied for high intensity laser pulses~\cite{CE5}.
 Therefore, another tool for the CEP determination for
 pulses with intensity $I\geq 10^{19}$ W/cm$^2$ needs to be
 developed. One avenue for getting access to the CEP
 is to search for observables which are both sensitive to
 the CEP variations and experimentally controllable.
 In Ref.~\cite{CE5}, the effect of CEP is demonstrated in the
 angular distribution of the photons emitted by a relativistic electron
 via multiphoton Compton scattering off an intense linearly polarized
 short pulse.
 The asymmetries of azimuthal photon distributions
 in nonlinear Compton scattering were analyzed in~\cite{CE8}.
 The impact of the CEP on  $\ee$ pair production with
 linearly polarized laser pulses is discussed in Ref.~\cite{CE6}.
 The importance of the CEP in Schwinger pair production in sub-cycle pulses
 is shown in~\cite{CE7}. {Some other aspects of the effect of CEP
 in $\ee$ pair productions are discussed recently in~\cite{A1,A2,A4,A5}.}

 Our present work may be considered as a further development of approaches
 to determine CEP for short and ultra-short {\it circularly polarized}
 laser pulses. We consider here both the generalized Breit-Wheeler process
 ($\ee$ pair production) and generalized Compton scattering (single photon emission
 off a relativistic electron).
 {We limit our discussion to circularly polarized pulses: First,
 because the case of linear laser polarization has been already
 analyzed in~\cite{CE5,CE6,CE7}. And second,
 the methods elaborated in \cite{TitovPRA,TitovEPJD} for
 quantum processes with circularly polarized finite laser pulses
 (by utilizing the generalized Bessel functions) allows to show the effect
 of the CEP on a transparent, almost qualitative level,
 which may be used as a powerful
 method for the CEP determination.}
 Thus, we show that
 the azimuthal angle distributions of the {emitted} electron (positron) in
 the first case, and the recoil electron ({emitted} photon) in the second case
 are in fact very sensitive to a CEP variation and can be used to fix the
 CEP value.

 Our paper is organized as follows.
 In Sect.~II we discuss the effect of the CEP
 in the case of the generalized Breit-Wheeler process.
 The impact of CEP on the generalized Compton scattering
 is analyzed in Sect.~III.
 Our summary is given in Sect.IV.

\section{effect of CEP in Breit-Wheeler $\mathbf{e^+e^-}$ pair production}

\subsection{The laser pulse}
As we mentioned above, here we consider the generalized Breit-Wheeler process,
 i.e.~the interaction of a probe photon $X$ with a laser beam $L$ in the reaction
 $X+L\to e^+ +e^-$, where a multitude of laser photons
 can participate simultaneously in the $\ee$ pair creation.
 The emphasis here is on short and intensive laser pulses.
 Long and weak laser pulses are dealt with in the standard
 textbook Breit-Wheeler process (cf.~the review paper~\cite{Ritus-79}).
 The different aspect of $\ee$ pair creation in finite pulses
 was also analyzed
 in Refs.~\cite{TitovPRA,TitovPRL,Nousch,Krajewska}.
 Below, we will concentrate mainly on the effect
 of  CEP, using the definitions of our Ref.~\cite{TitovPRA}

 In the following we use the widely employed
 the electromagnetic (e.m.) four-potential
 for a circularly polarized laser field in
 the axial gauge $A^\mu=(0,\,\mathbf{A}(\phi))$ with
\begin{eqnarray}
 \mathbf{A}(\phi)=f(\phi) \left( \mathbf{a}_1\cos(\phi+\tilde\phi)+ \mathbf
 {a}_2\sin(\phi+\tilde\phi)\right)~, \label{III1}
 \end{eqnarray}
 where $\tilde\phi$ is the CEP. The quantity
 $\phi=k\cdot x$ is the invariant phase with four-wave vector
 $k=(\omega, \mathbf{k})$, obeying the null field property $k^2=k\cdot
 k=0$ (a dot between four-vectors indicates the Lorentz scalar
 product) implying $\omega = \vert\mathbf{k}\vert$,
 $ \mathbf{a}_{(1,2)} \equiv \mathbf{a}_{(x,y)}$;
 $|\mathbf{a}_x|^2=|\mathbf{a}_y|^2 = a^2$, $\mathbf{a}_x \mathbf{a}_y=0$;
 transversality means $\mathbf{k} \mathbf{a}_{x,y}=0$ in the present gauge.
 The envelope function $f(\phi)$ with
 $\lim\limits_{\phi\to\pm\infty}f(\phi)=0$ accounts for the
 finite pulse duration. For simplicity and for the sake of numerical examples
 we use $f(\phi)$ in the form of a hyperbolic secant
 \begin{eqnarray}
 f(\phi)=\frac{1}{\cosh\frac{\phi}{\Delta}}~,
 \label{E1}
 \end{eqnarray}
 where the dimensionless quantity $\Delta$ is related to the pulse duration
 $2\Delta=2\pi N$, where $N$ has the meaning of a number of cycles in the pulse
 and it is related to the time duration
 of the pulse $\tau=2N/\omega$.
 {The carrier envelope phase $\tilde\phi$ is the main subject of our present discussion
 and, as we will show, its impact is particularly strong for short pulse duration $\Delta$.}

\subsection{Cross section}

 { The cross section of $\ee$ pair production is determined by the
 transition matrix $ M_{fi}(\ell)$ as
 \begin{eqnarray}
 \frac{d\sigma}{d\phi_e}=\frac{\alpha^2 v\,\zeta }{8m^4\xi^2N_0}\,
 \int\limits_{\zeta}^\infty d\ell \int\limits_{-1}^{1} d\cos\theta_e\,
 |M_{fi}(\ell,u)|^2~,\quad(\rm E1)
 \label{EQ1}
 \end{eqnarray}
 where  $m$ is the electron mass,
 $\theta_e$ is the polar angle
 of outgoing electron, $v$ is the electron velocity
 in c.m.s.
 In Eq.~(\ref{EQ1}) the averaging and sum over
 the spin variables in the initial and the final states
 is assumed; the azimuthal angle of the
 outgoing electron $\phi_e$, is defined as
 $\cos\phi_e={\mathbf a_x}{\mathbf p}_e/a |{\mathbf p}_e|$.
 The azimuthal angle of the positron momentum is
 $\phi_{e^+}=\phi_{e} + \pi$.
 The variable $\xi$ is the reduced field intensity $\xi^2=e^2a^2/m^2$.
 We use natural units with
 $c=\hbar=1$, $e^2/4\pi = \alpha \approx 1/137.036$.
 The lower limit of the integral is the threshold parameter $\zeta=4m^2/s$,
 where $s$ is the square of the total energy in the c.m.s.. The region of $\zeta<1$
 corresponds to the above-threshold $\ee$ pair production, while
 the region of $\zeta>1$ matches the sub-threshold pair production.
 Denoting four-vectors $k(\omega,{\mathbf k})$,
 $k'(\omega',{\mathbf k}')$,
 $p(E,{\mathbf p})$ and $p'(E',{\mathbf p}')$ as the four-momenta of the
 background (laser) field (\ref{III1}),
 incoming probe photon, outgoing positron and electron,
 respectively, the variables $s$, $v$ and $u$
 are determined as $s={2k\cdot k' }= 2(\omega'\omega -{\mathbf k}'{\mathbf k})$,
 $v^2=(\ell s-4m^2)/\ell s$,
 $u\equiv(k'\cdot k)^2/\left(4(k\cdot p)(k\cdot p')\right)=1/(1-v^2\cos^2\theta_e)$.
 The factor $N_0$ reads
$
 N_0={1}/{2\pi}\int\limits_{-\infty}^{\infty}
 d\phi\,(f^2(\phi)+ {f'}^2(\phi))
 $
 and determines the photon flux in case of the finite pulse~\cite{TitovEPJD}.
 The variable $\ell$ saves continuous values and the product
 $\ell\omega$ has the meaning of the laser energy
 involved into the process (see also~\cite{A1} for recent discussion).

 The transition matrix $ M_{fi}(\ell)$ in~(\ref{EQ1})
consists of four terms
\begin{eqnarray}
\, M_{fi}(\ell)=\sum\limits_{i=0}^3  M^{(i)}\,C^{(i)}(\ell)~,
\label{EQ2}
\end{eqnarray}
where the transition operators $ M^{(i)}$
read
\begin{eqnarray}
M^{(i)}=\bar u_{p'}\,\hat M^{BW(i)}\,v_p~
\label{B1}
\end{eqnarray}
with
\begin{eqnarray}
 \hat M^{BW(0)}&=&\fs\varepsilon'~,\quad
 \hat M^{BW(1)}=-
  \frac{ e^2a^2 \,
 (\varepsilon'\cdot k)\,\fs k}
 {2(k\cdot p)(k\cdot p')}~,\nonumber\\
 \hat M^{BW(2,3)}&=&\frac{e\fs a_{(1,2)}\fs k\fs
 \varepsilon'}{2(k\cdot p')} - \frac{e\fs \varepsilon'\fs k\fs
 a_{(1,2)}}{2(k\cdot p)}~,
 \label{B2}
\end{eqnarray}
where $u_{p'}$ and $v_{p}$ are the Dirac spinors of the electron and positron,
respectively, and
$\varepsilon'$ is the polarization four-vector of the probe photon.
The functions $C^{{i}}(\ell)$ can be written in terms
of the basic functions $Y_\ell$ and $X_\ell$ which may be considered as
the generalized Bessel functions for the finite e.m. pulse
\begin{eqnarray}
Y_\ell(z)&=&\frac{1}{2\pi} {\rm e}^{-i\ell(\phi_0-\tilde\phi)}\int\limits_{-\infty}^{\infty}\,
d\phi\,{f}(\phi)
\,{\rm e}^{i\ell\phi-i{\cal P}(\phi)} ~,\nonumber\\
X_\ell(z)&=&\frac{1}{2\pi}{\rm e}^{-i\ell(\phi_0-\tilde\phi)} \int\limits_{-\infty}^{\infty}\,
d\phi\,{f^2}(\phi)
\,{\rm e}^{i{\ell} \phi-i{\cal P}(\phi)}
\label{CP3}
\end{eqnarray}
 with
 \begin{eqnarray}
{\cal P(\phi)}&=&z\int\limits_{-\infty}^{\phi}\,d\phi'\,
\cos(\phi'-\phi_0+\tilde\phi)f(\phi')\nonumber\\
&-&\xi^2\zeta u\int\limits_{-\infty}^\phi\,d\phi'\,f^2(\phi')
\label{CP2}
\end{eqnarray}
 in the following form
 \begin{eqnarray}
C^{(0)}(\ell)&=&\widetilde Y_\ell(z){\rm e}^{i\ell(\phi_0-\tilde\phi)}~,\nonumber\\
\widetilde Y_\ell(z)&=&\frac{z}{2\ell} \left(Y_{\ell+1}(z) +
Y_{\ell-1}(z)\right) - \xi^2\frac{u}{u_\ell}\,X_\ell(z)~,\nonumber\\
C^{(1)}(\ell)&=&X_\ell(z)\,{\rm e}^{i\ell(\phi_0-\tilde\phi)}~,\nonumber\\
C^{(2)}({\ell})&=&\frac{1}{2}\left( Y_{\ell+1}{\rm e}^{i(\ell+1)\phi_0}
+ Y_{\ell-1}{\rm e}^{i(\ell-1)\phi_0}\right){\rm e}^{-i\ell\tilde\phi} ~,\nonumber\\
C^{(3)}({\ell})&=&\frac{1}{2i}\left( Y_{\ell+1}{\rm e}^{i(\ell+1)\phi_0}
- Y_{\ell-1}{\rm e}^{i(\ell-1)\phi_0}\right){\rm e}^{-i\ell\tilde\phi}~,\nonumber\\
\label{CP4}
\end{eqnarray}
 where the {argument $z$ of the generalized Bessel functions} is related to $\xi$, $\ell$, and $u$
 via
$
z=2{\ell}\xi\sqrt{{u}/{u_\ell}\left(1-{u}/{u_\ell}\right)}
$
with $u_\ell\equiv\ell/\zeta$ and $\phi_0$ is the azimuthal angle of outgoing electron
$\phi_0=\phi_e$.
 Eq.~(\ref{CP4}) allows to express the differential cross section
in terms of the partial probabilities $w(\ell)$
\begin{eqnarray}
\frac{d \sigma}{d\phi_e }
=\frac{\alpha^2v\,\zeta}{4m^2\xi^2 N_0}
\,\int\limits_\zeta^{\infty}\,d\ell
\int\limits_{-1}^{1} d\cos\theta_e
\, w{(\ell)}~
\label{III99}
\end{eqnarray}
with
\begin{eqnarray}
w(\ell)&=&
 2 |\widetilde Y_\ell(z)|^2+\xi^2(2u-1)\nonumber\\
&\times&\left(|Y_{\ell-1}(z)|^2 + |Y_{\ell+1}(z)|^2 -2
{\rm Re}\,(\widetilde Y_\ell(z)X^*_\ell(z))\right)~,\nonumber\\
 \label{III26-0}
\end{eqnarray}
which resembles the known expression for
the partial probability in case of
the infinitely long e.m. pulse~\cite{Ritus-79}
\begin{eqnarray}
w_n=
 2 J^2_n(z') +\xi^2(2u-1)
\left(J^2_{n-1}(z')+ J^2_{n+1}(z')-2J^2_n(z')\right)
\nonumber
\end{eqnarray}
with the substitutions $\ell\to n$, $|\widetilde Y_\ell(z)|^2\to J^2_n(z')$,
$|Y_{\ell\pm1}(z)|^2\to J^2_{n\pm1}(z')$,
${\rm Re}\,(\widetilde Y_\ell(z)X^*_\ell(z))\to J^2_n(z')$,
and
$
z'=(2n\xi)/(1+\xi^2)^{1/2}\sqrt{{u}/{u_n}\left(1-{u}/{u_n}\right)}
$
with $u_n\equiv n/\zeta$.

\subsection{Numerical results}

 It is naturally to expect that the  effect of the finite carrier
 phase essentially appears in the azimuthal angle distribution of
 the outgoing electron (positron)
 because the carrier phase is included in the expressions for the basic
 functions (\ref{CP3}) in the combination $\phi_e-\tilde\phi$.
  As an example, in Fig.~\ref{Fig:CE1} (left panels) we show
  the differential cross section $d\sigma/d\phi_e$
 of $\ee$ pair production as a function of the azimuthal angle
 $\phi_e$ for different values of the carrier envelope phase
 $\tilde\phi$
 and for different pulse durations
 $\Delta=N\pi$ with $N=$ 0.5, 1, 1.5 and 2, and for $\xi^2=0.5$.
 The calculation is done for the essentially multi-photon region with
 $\zeta=4$.

 \begin{figure}[t]
\includegraphics[width=0.45\columnwidth]{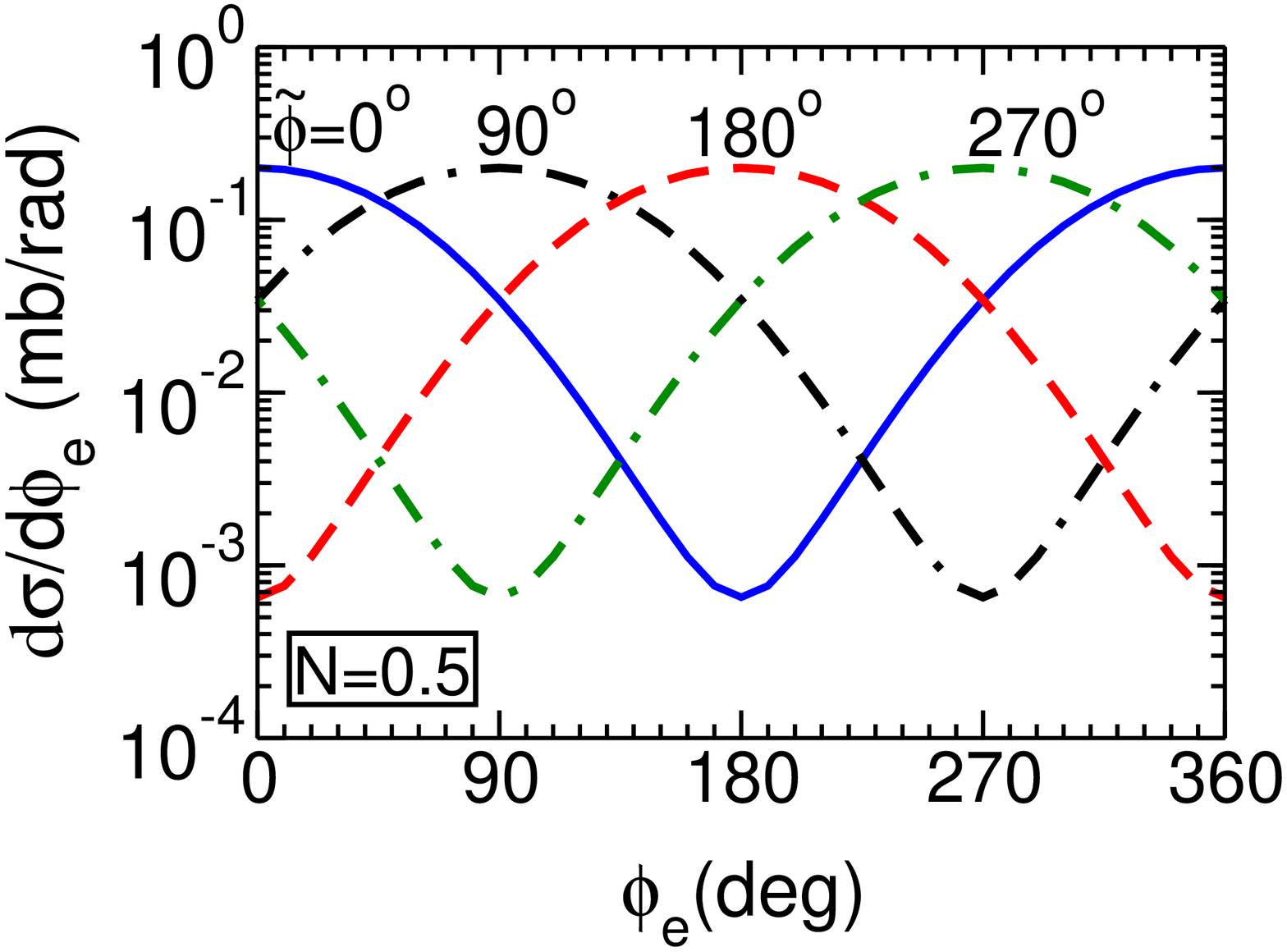}\qquad
\includegraphics[width=0.45\columnwidth]{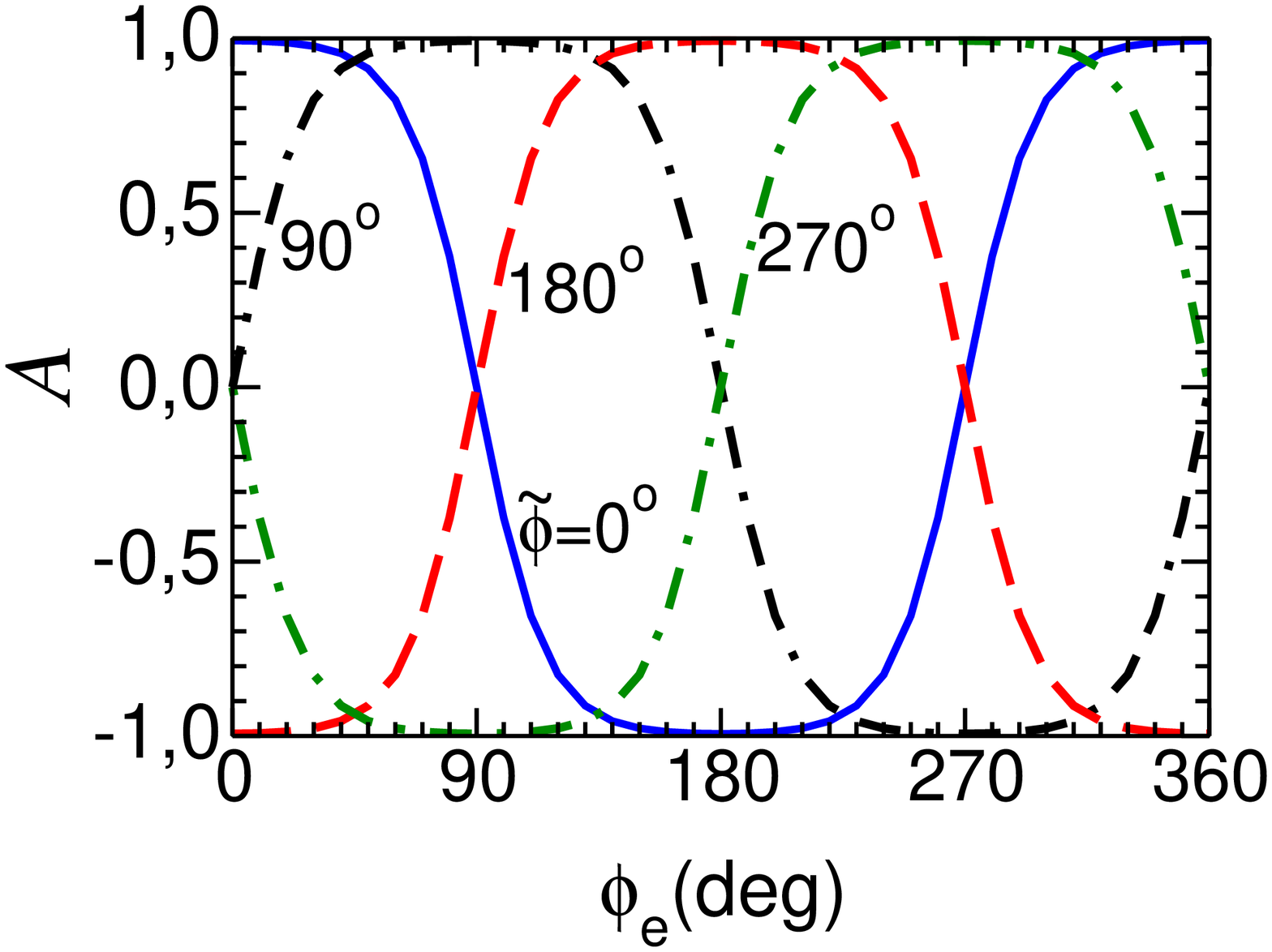}
\includegraphics[width=0.45\columnwidth]{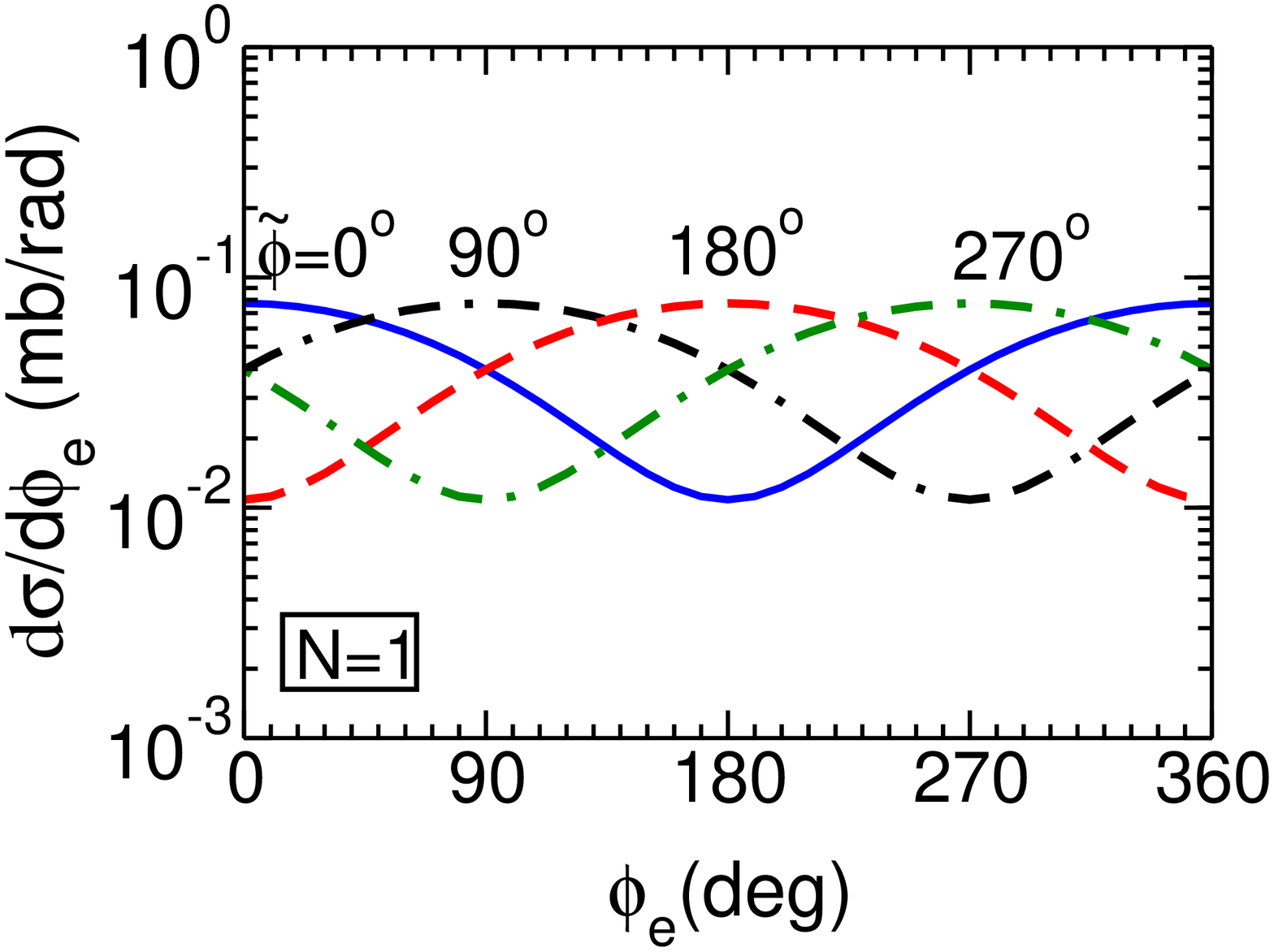}\qquad
\includegraphics[width=0.45\columnwidth]{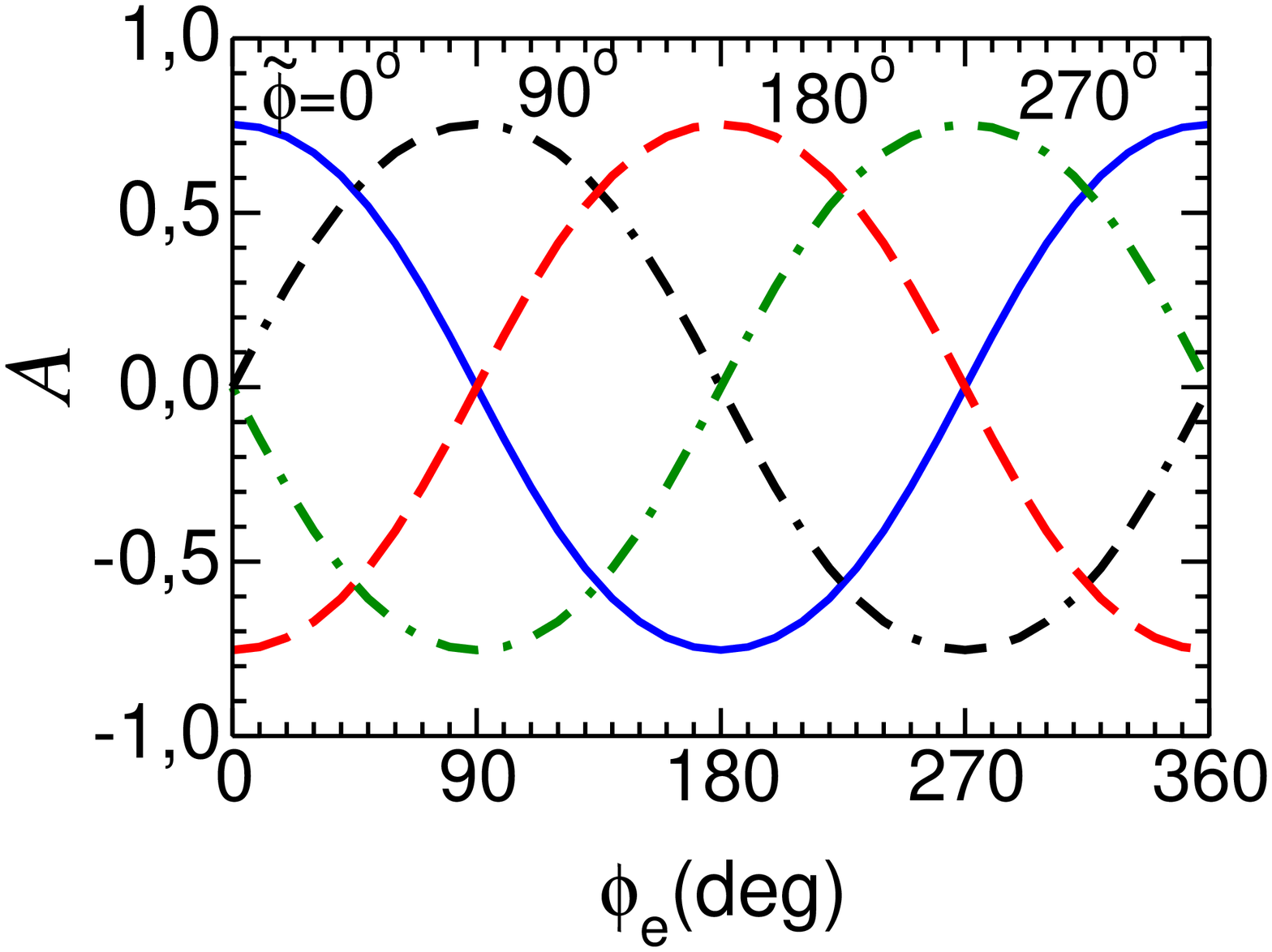}
\includegraphics[width=0.45\columnwidth]{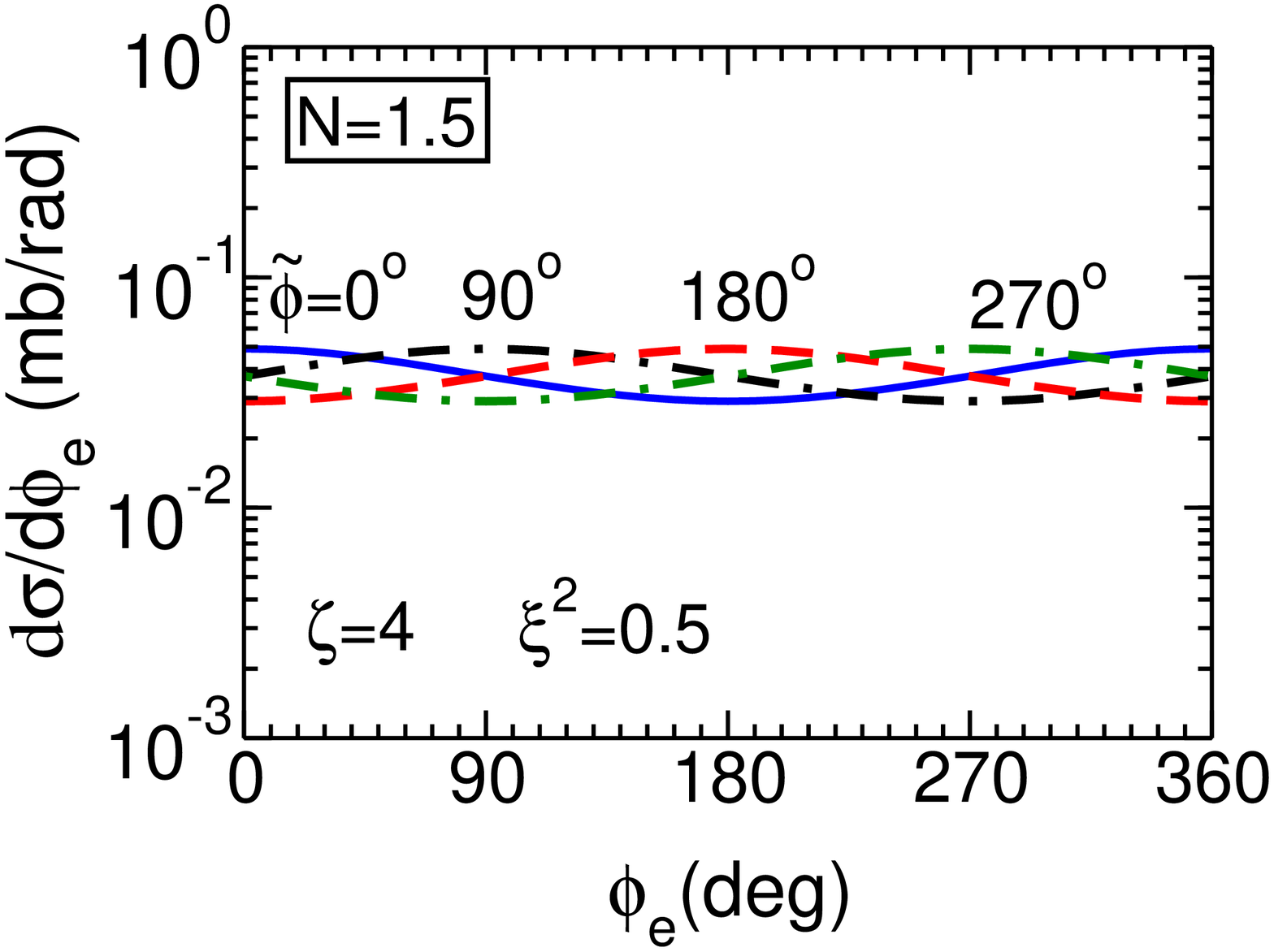}\qquad
\includegraphics[width=0.45\columnwidth]{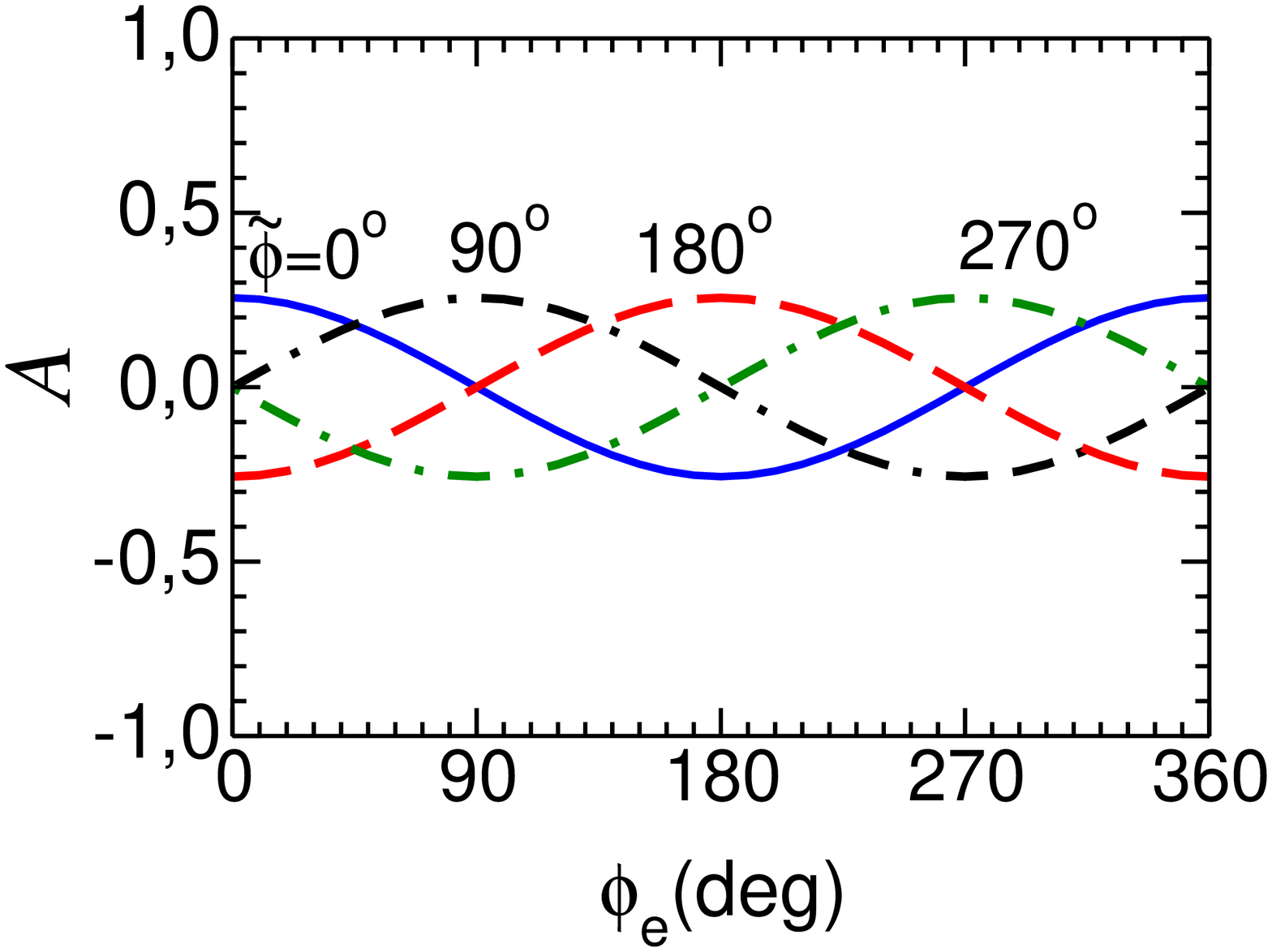}
\includegraphics[width=0.45\columnwidth]{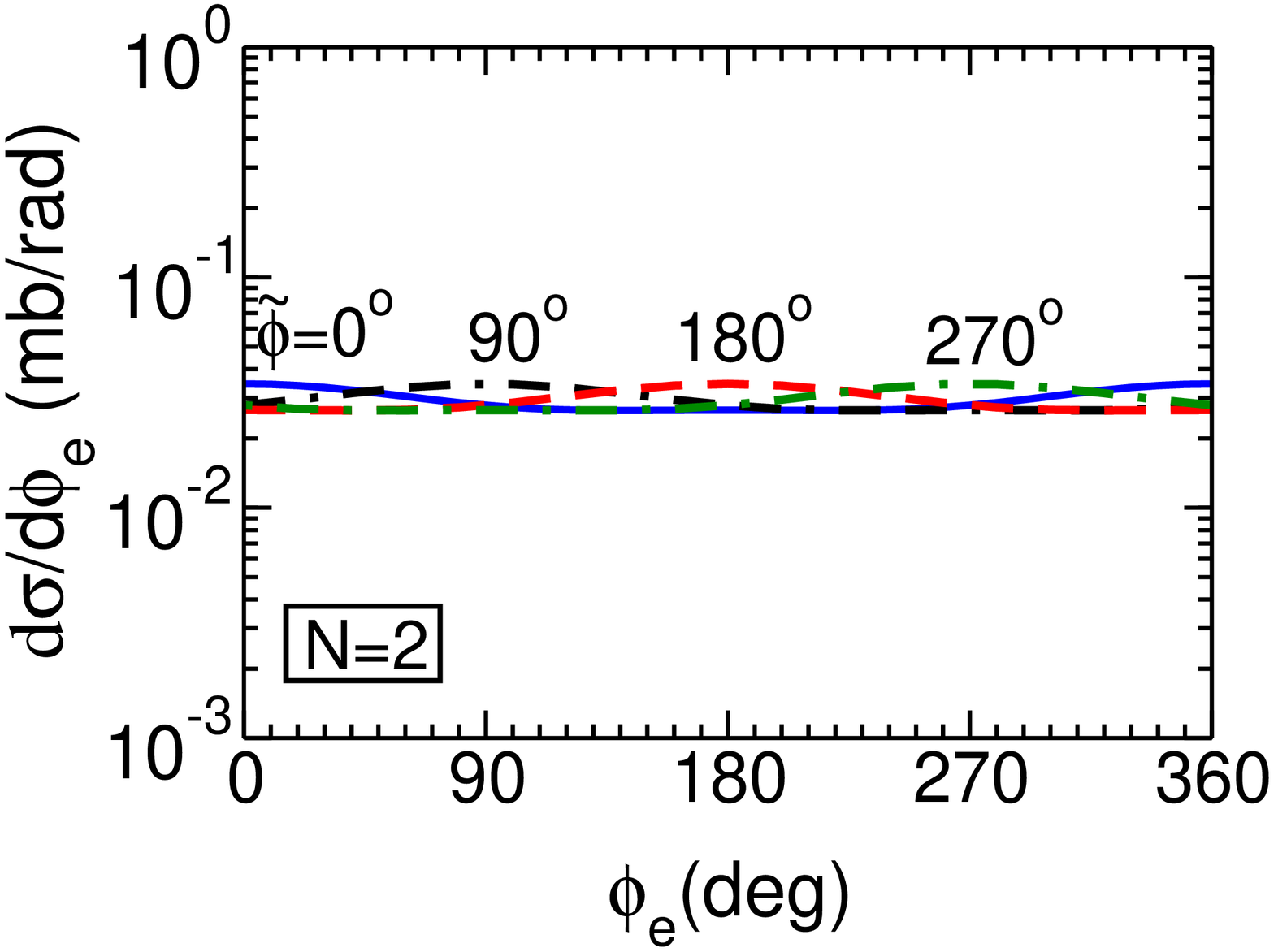}\qquad
\includegraphics[width=0.45\columnwidth]{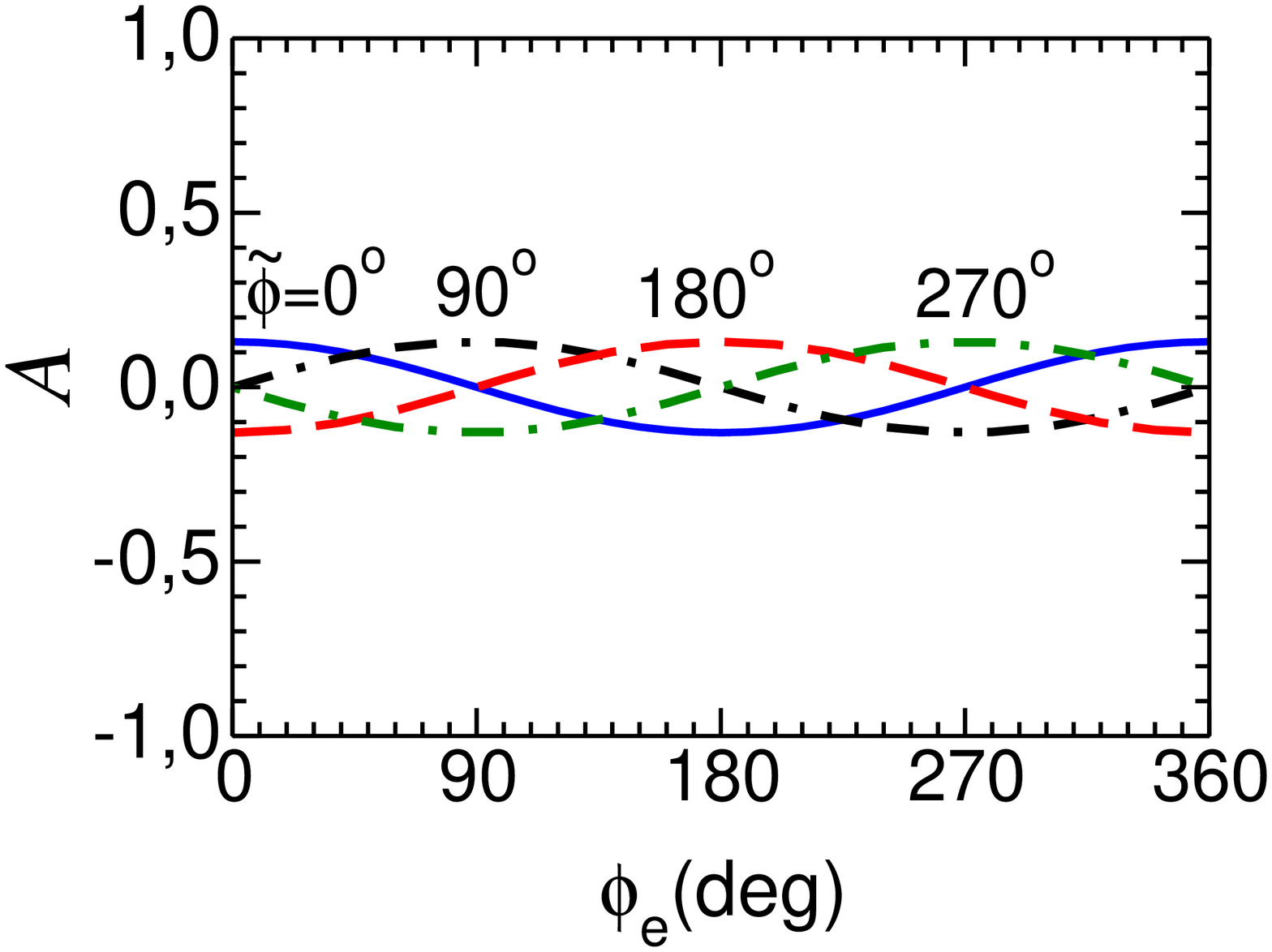}
\caption{\small{(Color online)
Left column: The differential cross section as a function of the azimuthal
angle of the direction of flight of the outgoing electron
$\phi_e$, for different values of the
carrier phase $\tilde\phi$ and for different pulse duration
$\Delta=N\pi$ with $N=$0.5, 1, 1.5 and 2 (from the bottom panels).
The solid, dash-dash-dotted, dashed and dash-dotted
curves are for the CEP equal to 0,
90, 180 and 270 degrees, respectively.
Right column: The anisotropy~(\ref{U9}) for different
values of $\tilde\phi$ and $N$, as in left column.
For $\xi^2=0.5$ and $ \zeta=4$.
 \label{Fig:CE1}}}
\end{figure}

 One can see a clear bump-like structure of the cross sections,
 in particularly for very short pulses with $N\leq1$,
 where the bump position coincides with the corresponding value of
 the carrier phase. In these cases  the height of the bumps
 can reach orders of magnitude.
 The reason of such behaviour {is the following:}
 The basic functions $Y_\ell$ and $X_\ell$ are determined by the integral
 over $d\phi$ with a rapidly oscillating exponential function $\exp[i\Psi]$ with
\begin{eqnarray}
\Psi= \ell\phi
 -&z&(\cos(\phi_e-\tilde\phi)\int\limits_{-\infty}^{\phi} d\phi'\, f(\phi')\cos\phi'
 \nonumber\\
 &+& \sin(\phi_e-\tilde\phi)\int\limits_{-\infty}^{\phi} d\phi'\, f(\phi')\sin\phi'
 ).
 \label{UU8}
 \end{eqnarray}
 Then, taking into account the inequality for $\phi>0$
 \begin{eqnarray}
 \int\limits_{-\infty}^{\phi} d\phi'\, f(\phi')\cos\phi'
 \gg
 \int\limits_{-\infty}^{\phi} d\phi'\, f(\phi')\sin\phi'~,
 \label{UU81}
\end{eqnarray}
 which is valid for the smooth sub-cycle pulse shapes
 (e.g.~for the hyperbolic secant shape), one can conclude
 that the main contribution to the probability comes from the region
 $\phi_e\simeq\tilde\phi$, which is confirmed by the result of our full
 calculation shown in Fig.~\ref{Fig:CE1} (left panels).
 The effect of the carrier phase decreases with increasing pulse
 duration and for $N\ge 2$ it becomes negligibly small.

  The corresponding anisotropies of the electron (positron) emission defined
  as
  \begin{eqnarray}
{\cal A}=\frac{d\sigma(\phi_e) - d\sigma(\phi_e+\pi)}
{d\sigma(\phi_e) + d\sigma(\phi_e+\pi)}~,
\label{U9}
\end{eqnarray}
  are exhibited in
  Fig.~\ref{Fig:CE1} (right panels). One can see a strong dependence
  of the anisotropy on CEP, especially for the sub-cycle pulses
  with $N=0.5$, 1 which is consequence of the
  "bump"\ structure of the differential cross sections shown in the
  left panels.
  In these cases the anisotropy takes a maximum value ${\cal A}\simeq 1$
  at $\phi_e=\tilde\phi$ and  $|{\cal A}|<1$ at $\phi_e\neq\tilde\phi$.
  It takes a minimum value ${\cal A}\simeq -1$
  at  $\phi_e-\tilde\phi=\pm\pi$. The increase of the pulse duration leads
  to a decrease of the absolute value of ${\cal A}$

 Since the basic functions in (\ref{III26-0}) depend on the CEP
 through ${\cal P}(\phi)$ (cf. (\ref{CP2})) solely in combination
 $\phi_e-\tilde\phi$, then the cross sections and
 anisotropies depend on the scale variable $\Phi=\phi_e-\tilde\phi$.
 This means that the dependence of
 observables as a function of the {re-scaled azimuthal angle} $\Phi$
 are independent of the CE phase $\tilde\phi$ and they coincide
 with the dependence of these observables on $\phi_e$ at $\tilde\phi=0$.

 \begin{figure}[t]
\includegraphics[width=0.45\columnwidth]{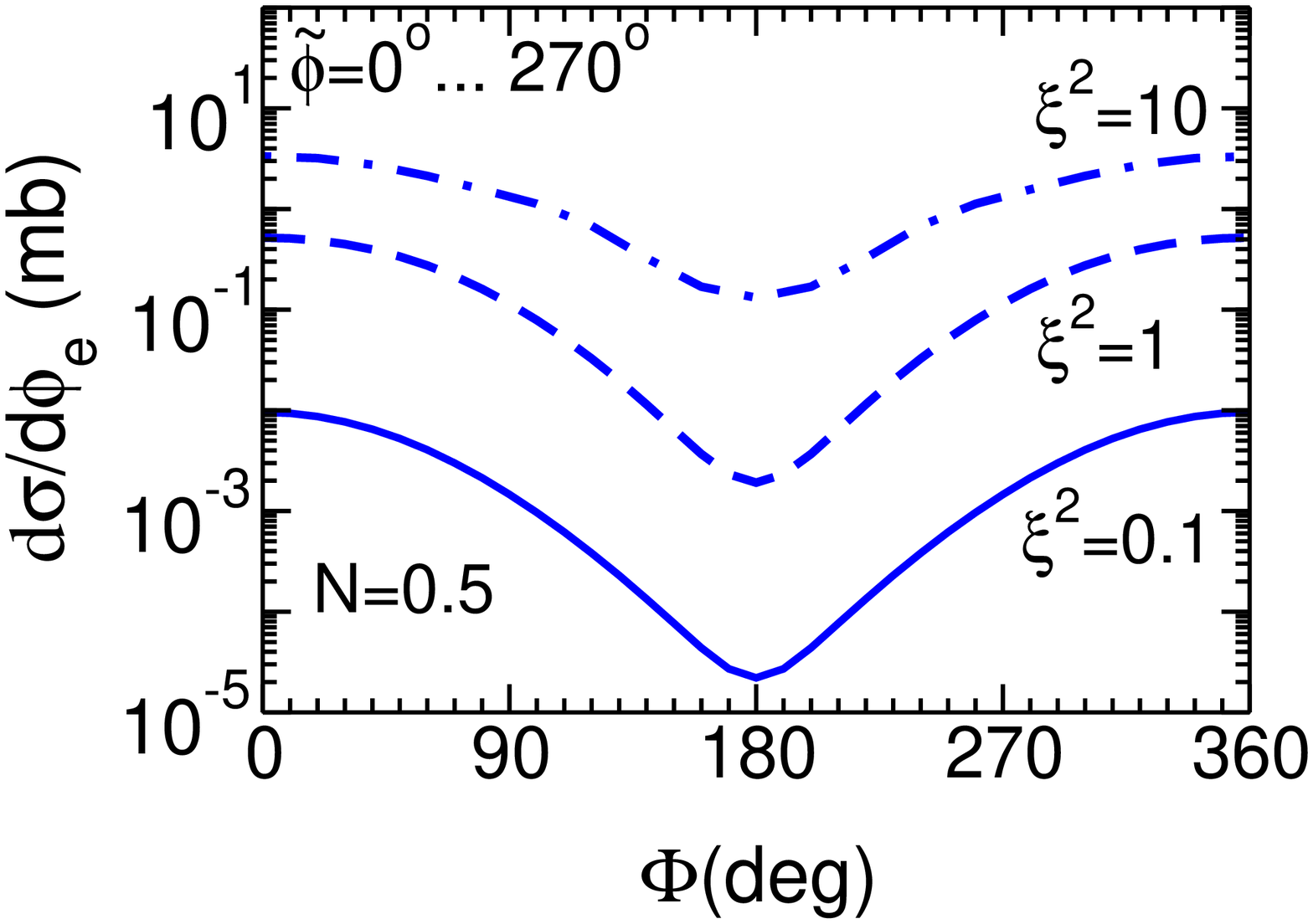}\qquad
\includegraphics[width=0.45\columnwidth]{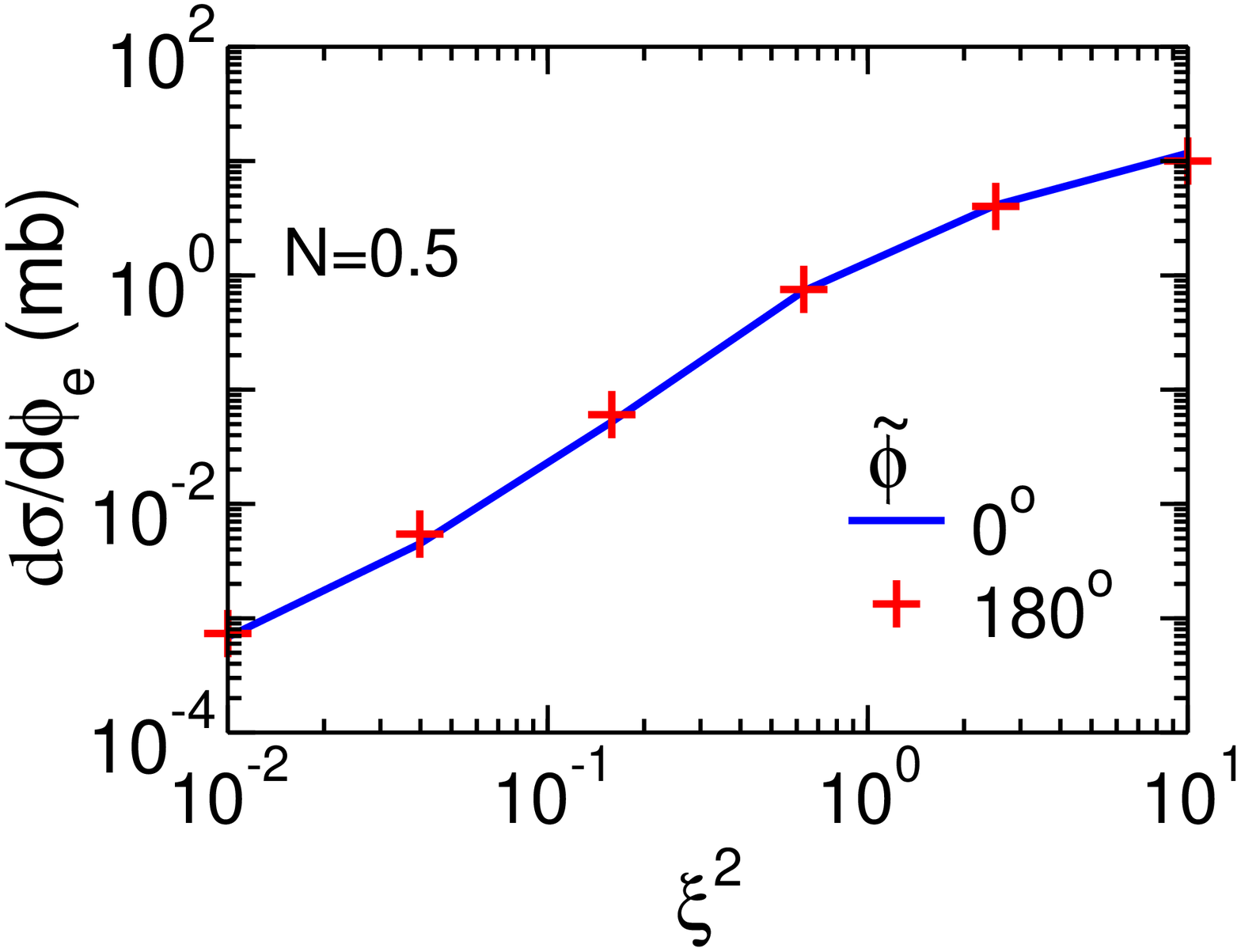}
\caption{\small{(Color online)
Left panel: The cross section of the $\ee$ production as a function of
scale variable $\Phi=\phi_e-\tilde\phi$ for different values
of e.m. field intensities $\xi$.
The solid, dashed and dash-dotted
curves are for $\xi^2=$0.1, 1, and 10, respectively.
Right panel:
The cross section of $\ee$ pair production integrated
over $\phi_e$
as a function of $\xi^2$ for $\tilde\phi=0$ (solid curve)
and $\pi$ (crosses).
For $N=0.5$ and $\zeta=4$.
 \label{Fig:CE2}}}
\end{figure}

 In Fig.~\ref{Fig:CE2} (left panel) we show the cross sections
 of $\ee$ pair production
 as a function of $\Phi$ for the sub-cycle pulse with $N=0.5$
 at different values of field intensity $\xi^2=$0.1, 1, and 10.
 One can see, that qualitatively the shape of the cross section
 is not  sensitive to the value of $\xi^2$. However the height
 of the bumps in the cross section slightly decreases with increasing
 $\xi^2$.

 Finally we note that the total cross section (integrated over $\phi_e$)
 is not sensitive to the CEP. Thus, Fig.~\ref{Fig:CE2} (right panel)
 exhibits the total cross section of the $\ee$ pair production
 for the sub-cycle pulse with $N=0.5$ and $\zeta=4$ as a function of $\xi^2$.
 The solid curve and crosses correspond to
 $\tilde\phi=0$ and $\pi$, respectively. One can see in fact
 that the total cross section is independent of CEP.


\section{Impact of CEP on the generalized Compton scattering}

\subsection{The cross section}

 The Compton scattering process, symbolically
 $e^-+ L\to e^{-}{}' + \gamma' $ is considered here
 as the spontaneous emission
 of one photon off an electron in an external e.m.\ field (\ref{III1}).
 Some important aspects of generalized Compton scattering
 were discussed elsewhere, see, for
 instance~\cite{CE8,Boca-2009,Heinzl-2009,Seipt-2011,Dinu,Seipt-2012,Krajewska-2012},
 and recent~\cite{A3}.
 Below we use the notations and definitions of Ref.~\cite{TitovEPJD}.

 We denote the four-momenta of the
 incoming electron, background (laser) field (\ref{III1}),
 outgoing electron and photon as, $p(E,{\mathbf p})$, $k(\omega,{\mathbf k})$,
 $p'(E',{\mathbf p}')$, $k'(\omega',{\mathbf k}')$,
 respectively.

 The cross section of the Compton scattering is determined by the
 transition matrix $ M_{fi}(\ell)$ as
 \begin{eqnarray}
 \frac{d\sigma}{d\omega'd\phi_{e'}}=\frac{\alpha^2}{N_0\xi^2(s-m^2)m^2}\,
 \int\limits_{\delta \ell}^\infty d\ell
 \frac{|M_{fi}(\ell)|^2}{||\mathbf p|-\ell\omega|}~,
 \label{CPC1}
 \end{eqnarray}
 where the averaging and sum over the spin variable in the initial and the final states
 is assumed. The lower limit $\delta\ell$ may be smaller than 1
 and, for instance, it may be determined by
 the detector resolution $\delta\omega'$ according to
 relation between
  the frequency $\omega'$ of the emitted photon,
 auxiliary variable $\ell$  and the polar angle $\theta'$
 of the direction of the momentum $\mathbf{k}'$ thought the
 conservation laws as
\begin{eqnarray}
 \omega'=\frac{\ell\,\omega (E+|\mathbf {p}|)}{E + |\mathbf {p}| \cos\theta'
 +\ell \omega(1-\cos\theta') }~.
 \label{CPC7}
\end{eqnarray}
 The frequency $\omega'$ increases with $\ell$ at fixed $\theta'$ since $\omega'$ is
 a function of $\ell$ at fixed $\theta'$.
 All quantities are considered in the laboratory system.

 Being a crossing channel to the Breit-Wheeler process, the main features of the
 Compton scattering formalism and the results
 are very close to that of the Breit-Wheeler process,
 although there are some differences. Thus, there is
 no evident sub-threshold effect because all frequencies
 $\omega'$ can contribute
 in (\ref{CPC1}) (see, however, below we define specific
 subthreshold parameter responsible for the multi-photon effects).

  The transition matrix $M(\ell)$ consists of four terms,
 \begin{eqnarray}
 M(\ell)=\sum\limits_{i=0}^3  M^{C(i)}\,C^{(i)}(\ell)~,
 \label{CPC2}
 \end{eqnarray}
 where the transition operators
 $M^{C(i)}$ are related to the transition operator
 $M^{BW(i)}$ in (\ref{B2}) as $M^{C(i)}(p,p',k,k')=M^{BW(i)}(-p,p',k,-k')$.
 {The squared c.m.s.~energy reads $s=m^2+2k\cdot p$.}
 The coefficient functions $C^{(i)}(\ell)$ are determined by the basic
 functions according to Eqs.~(\ref{CP4}) but with the own
 phase function
 \begin{eqnarray}
  {\cal P(\phi)} &=& z\int_{-\infty}^{\phi}\,d\phi'\,
 \cos(\phi'-\phi_{e'}+\tilde\phi)f(\phi')
 \nonumber\\
 &-&\xi^2\frac{u}{u_{1}}
 \int_{-\infty}^\phi\,d\phi'\,f^2(\phi')~,
 \label{CPC3}
\end{eqnarray}
 where the azimuthal angle of the recoil electron
 $\phi_{e'}$ coincides with the angle $\phi_0$ in
 Ritus notation~\cite{Ritus-79,LL4} and is
 determined as $\cos\phi_{e'}={\mathbf a_x}{\mathbf p'}/a|{\mathbf p'}|$.
 The azimuthal angle of the photon momentum is
 $\phi_{\gamma'}=\phi_{e'} + \pi$.
 For the variables in Eq.~(\ref{CPC3})
 we use the standard notation:
 $z=2\ell\xi\left(({u}/{u_\ell})(1-{u}/{u_\ell})\right)^{1/2}$ with
 $u\equiv(k'\cdot k)/(k\cdot p')$, $u_\ell=\ell\,u_{1}$ and
 $u_{1} ={ (s-m^2)/m^2} = 2{k\cdot p}/m^2$.
 This representation of functions $C^{(i)}(\ell)$ allows to define the
 differential cross section through the partial probabilities
 $w(\ell)$
 \begin{eqnarray}
 \frac{d\sigma}{d\omega'd\phi_{e'}}
 =\frac{2\alpha^2}{N_0\,\xi^2\,(s-m^2)}\,
 \int\limits_{\delta \ell}^\infty d\ell\frac{w(\ell)}{||{\mathbf p}| - \ell\omega|}
  \label{CPC4}
 \end{eqnarray}
with
\begin{eqnarray}
&& w(\ell)=
 -2 |\widetilde Y_\ell(z)|+\xi^2\left( 1 +\frac{u^2}{2(1+u)}\right)
 \nonumber\\
 &&\times \left(|Y_{\ell-1}(z)|^2+ |Y_{\ell+1}(z)|^2
 -2{\rm Re}\,(\widetilde Y_\ell(z)X^*_\ell(z))\right)~.
 \label{CPC5}
\end{eqnarray}
 Equation (\ref{CPC5}) resembles the corresponding expression for the partial
 probability of photon emission in the case
 of the infinitely long pulse~\cite{Ritus-79,LL4}
 with the substitutions $\ell\to n$, $|\widetilde Y_\ell(z)|^2\to J^2_n(z')$,
$|Y_{\ell\pm1}(z)|^2\to J^2_{n\pm1}(z')$, and
${\rm Re}\,(\widetilde Y_\ell(z)X^*_\ell(z))\to J^2_n(z')$,
 \begin{eqnarray}
 w_n&=&
 -2 J^2_n(z')+\xi^2 \left(1 +\frac{u^2}{2(1+u)} \right)
 \nonumber\\
 &\times& \left(J^2_{n-1}(z')+ J^2_{n+1}(z')
 -2 J_n^2(z')\right)~,
 \nonumber
\end{eqnarray}
 where $J_n(z')$ denotes Bessel functions with
  $z'={2n\xi}/{({1+\xi^2})^{1/2}}({u}/{u_n}\left(1-{u}/{u_n}\right))^{1/2}$
  and  $u_n=(2n(k\cdot p))/(m^2(1+\xi^2))$.

 We recall that the internal quantity $\ell$ is a continuous
 variable, implying a continuous distribution
 of the differential
 cross section over the $\omega' - \theta'$ plane.
 The case of  $\ell = 1$ with $\xi^2\ll 1$
 and $\omega'_1=\omega'(\ell=1)$ recovers
 the Klein-Nishina cross section, cf.~\cite{Ritus-79}.
 The quantity $\ell\omega$ can be considered as an energy
 of the laser beam involved in the Compton process,
 which is not a multiple $\omega$.
 Mindful of this fact, without loss of generality, we
 denote the processes with $\ell>1$ as a multi-photon
 generalized Compton scattering, remembering
 that $\ell$ is a continuous quantity.
  We stress again, the internal
 variable $\ell$ can not be interpreted strictly as number of laser
 photons involved (cf.~\cite{DSeipt-2014}).

 The cross section of the multi-photon Compton scattering (for $\ell >
 1$) first increases with increasing $\theta'$ and
 peaks at a scattering angle $\theta'<180^\circ$, beyond which it rapidly
 drops to zero when $\theta'$ approaches $180^\circ$,
 yielding the blind spot in the high harmonics for back-scattering. For
 instance, the cross section
 peaks at about $170^\circ$ for the chosen electron energy of 4 MeV (all
 quantities are considered in the laboratory frame).

 Therefore, in our subsequent analysis we choose the near-backward photon
 production at $\theta' = 170^\circ$
 and an optical laser with $\omega=1.55$~eV. Defining
 one-photon events by $\ell=n = 1$, this kinematics leads via
 Eq.~(\ref{CPC7}) to $\omega_1'\equiv \omega'(l=1,
 \theta' = 170^o) \simeq 0.133$~keV which we refer as a
 threshold value. Accordingly, $\omega' > \omega'_1$ is enabled by
 non-linear effects which in turn may be related loosely to
 multi-photon dynamics.

 The ratio $\kappa=\omega'/\omega'_1$ may be considered as
 a sub-threshold parameter (as an analog of sub-threshold parameter
 $\zeta$ in the Breit-Wheeler process).
 Using the value $\kappa\simeq\ell'(\omega'(\ell'))$
  we can define the sub-threshold part the total cross section
 $\tilde\sigma$ explicitly
 as integral
 \begin{eqnarray}
 \frac{\tilde\sigma(\omega')}{d\phi_{e'}} = \int\limits_{\omega'}^{\infty}
 d\bar\omega \frac{d\sigma (\bar\omega)}{d\bar\omega d\phi_{e'}}
 =\int\limits_{\ell'(\kappa)}^{\infty} d\ell
 \frac{d\sigma(\ell)}{d\ell d\phi_{e'}}~,
 \label{S6}
\end{eqnarray}
 where
 $d\sigma(\ell) / d\ell d\phi_{e'} =
 ( d\sigma(\omega') / d\omega'd\phi_{e'})
 (d\omega'(\ell) / d\ell)$,
 and the minimum value of $\ell'(\kappa)$ is
 \begin{eqnarray}
 \ell'(\kappa)=\kappa
 \frac{E+|\mathbf p|\cos\theta'}{E+|\mathbf p|\cos\theta'-\omega(\kappa-1)(1-\cos\theta')}~.
 \label{S66}
 \end{eqnarray}
 The cross section (\ref{S6}) has the meaning of a cumulative distribution.
 In this case, the subthreshold, multi-photon events correspond to
 frequencies $\omega'$ of the outgoing photon which
 exceed the corresponding
 threshold value $\omega_1'=\omega'(\ell=1)$ (cf. Eq.~(\ref{CPC7})).

 Similarly to the Breit-Wheeler process, effect of the finite CEP
 essentially appears in the differential cross sections
 as a function of the
 azimuthal angle of the outgoing electron (photon) momentum
 because the carrier phase is included in the expressions for the basic
 functions (\ref{CP3}) with (\ref{CPC3})
 in combination $\phi_{e'}-\tilde\phi$.

\subsection{Numerical results}

\begin{figure}[ht]
\includegraphics[width=0.45\columnwidth]{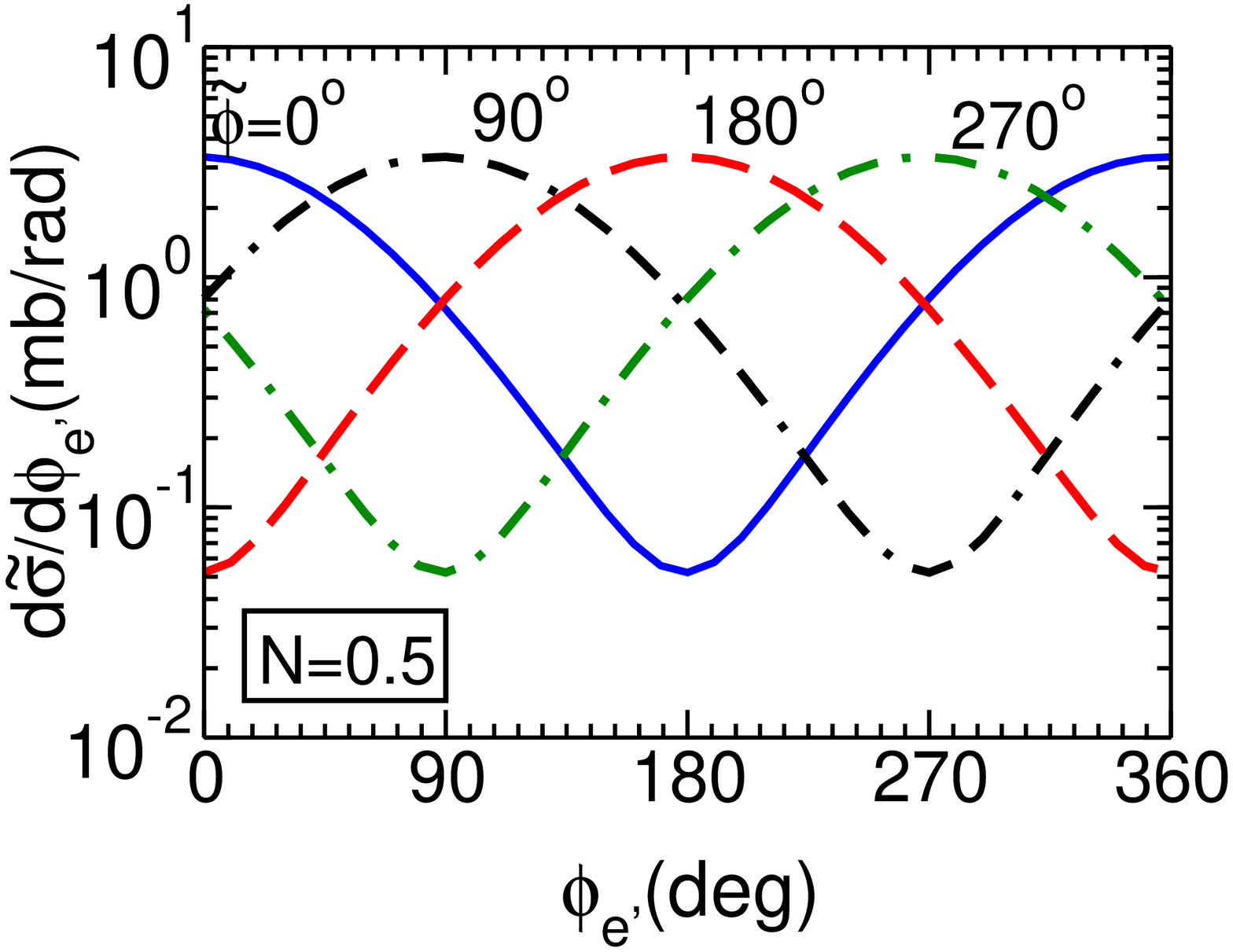}\qquad
\includegraphics[width=0.45\columnwidth]{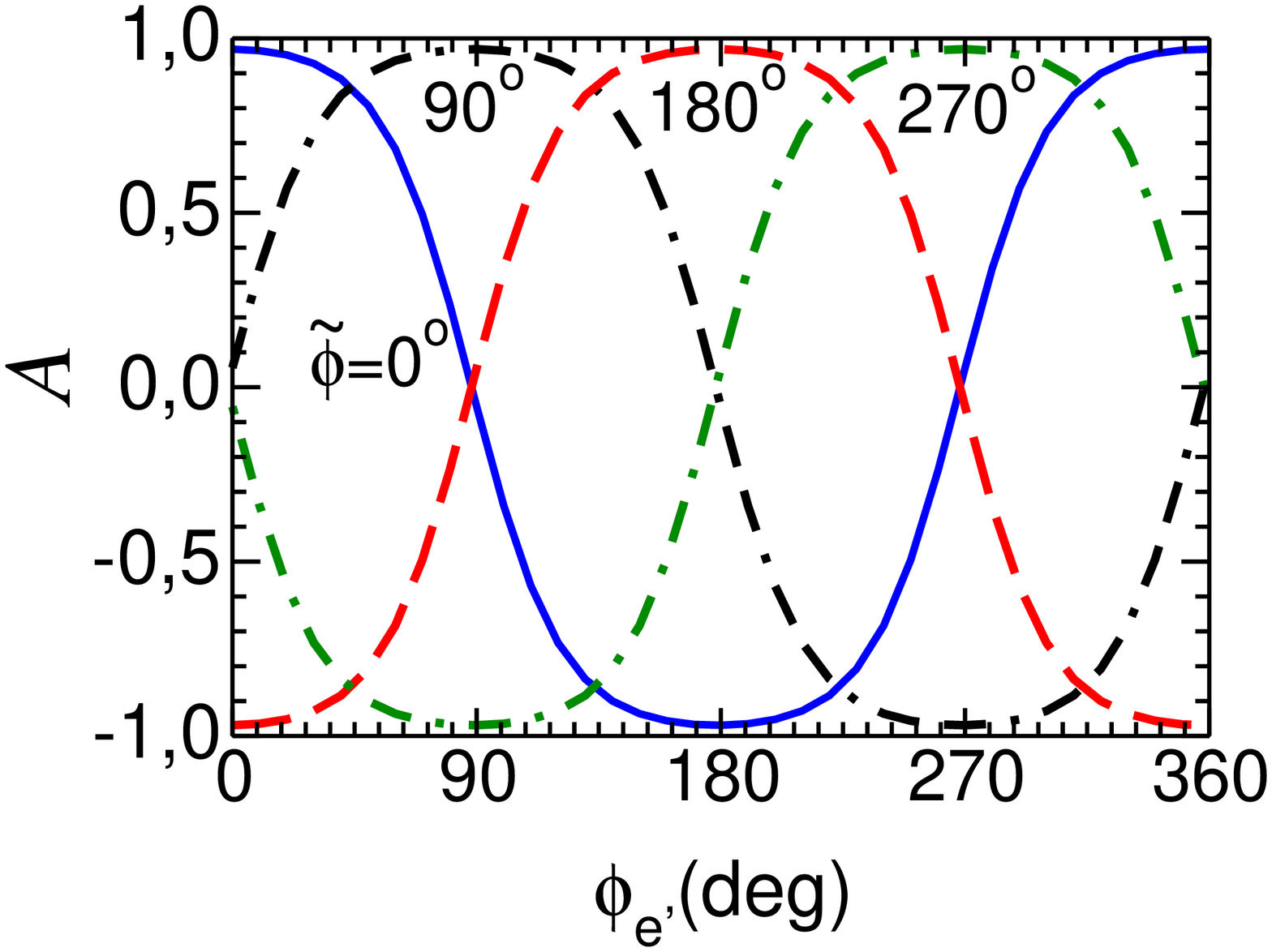}
\includegraphics[width=0.45\columnwidth]{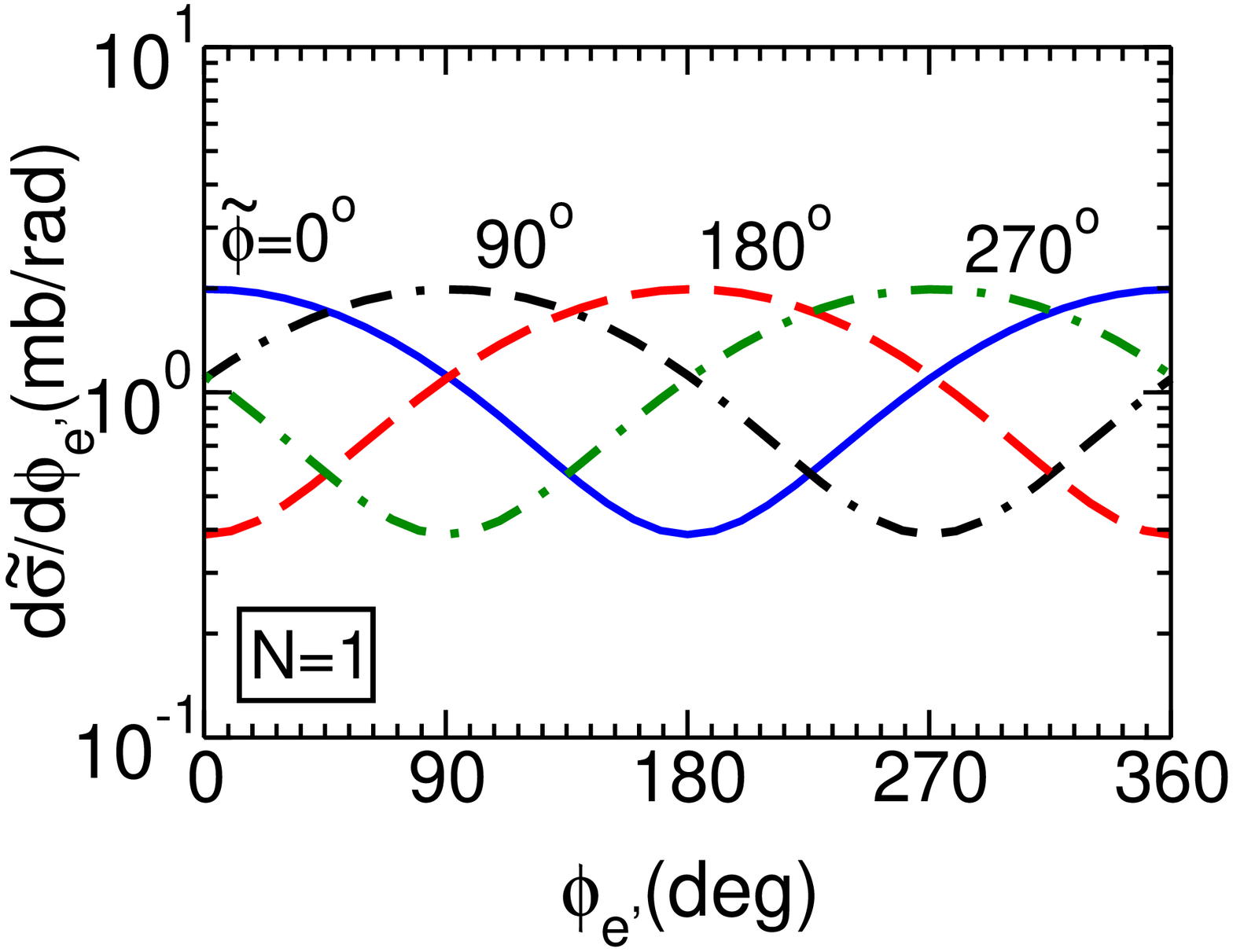}\qquad
\includegraphics[width=0.45\columnwidth]{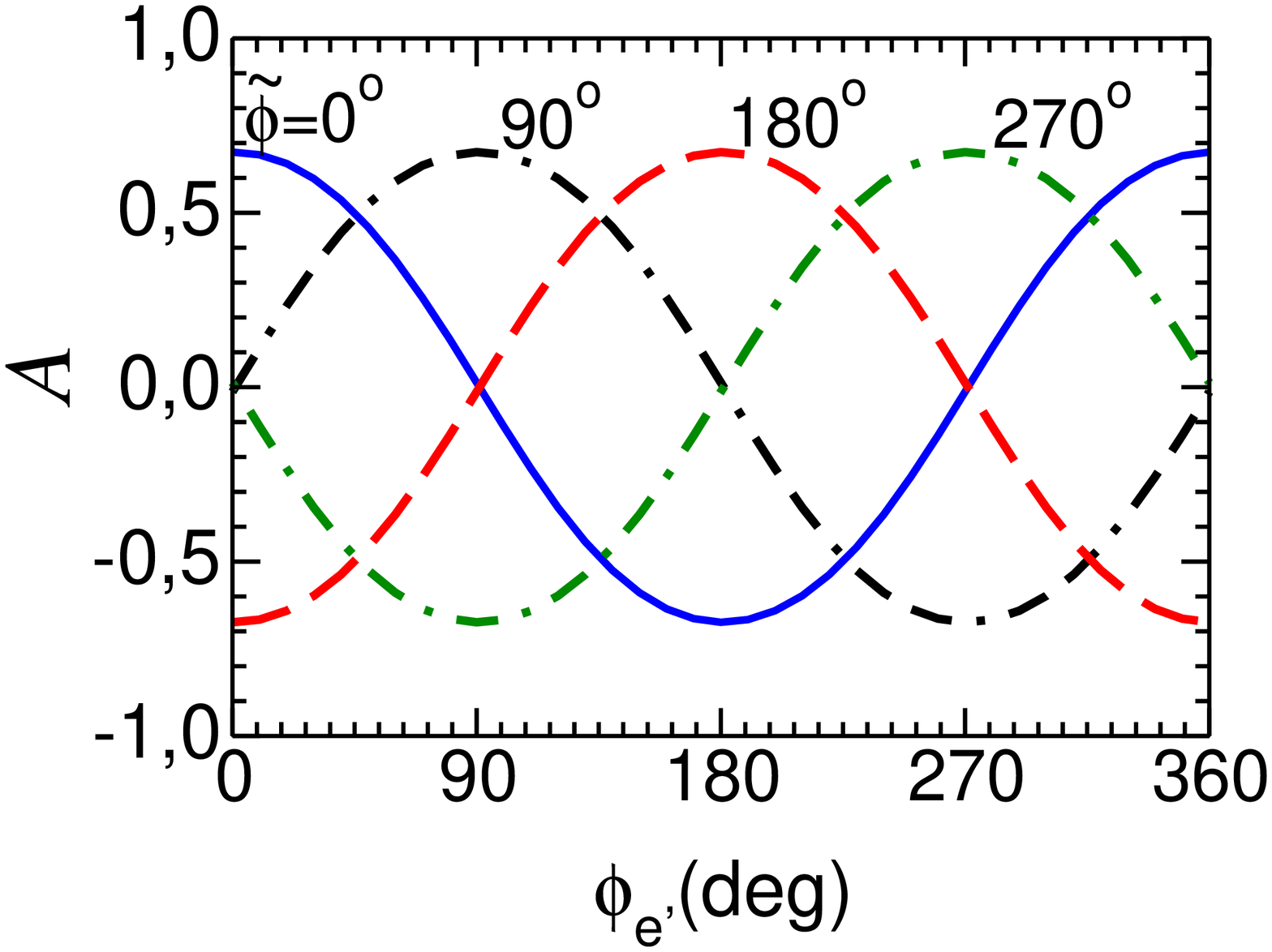}
\includegraphics[width=0.45\columnwidth]{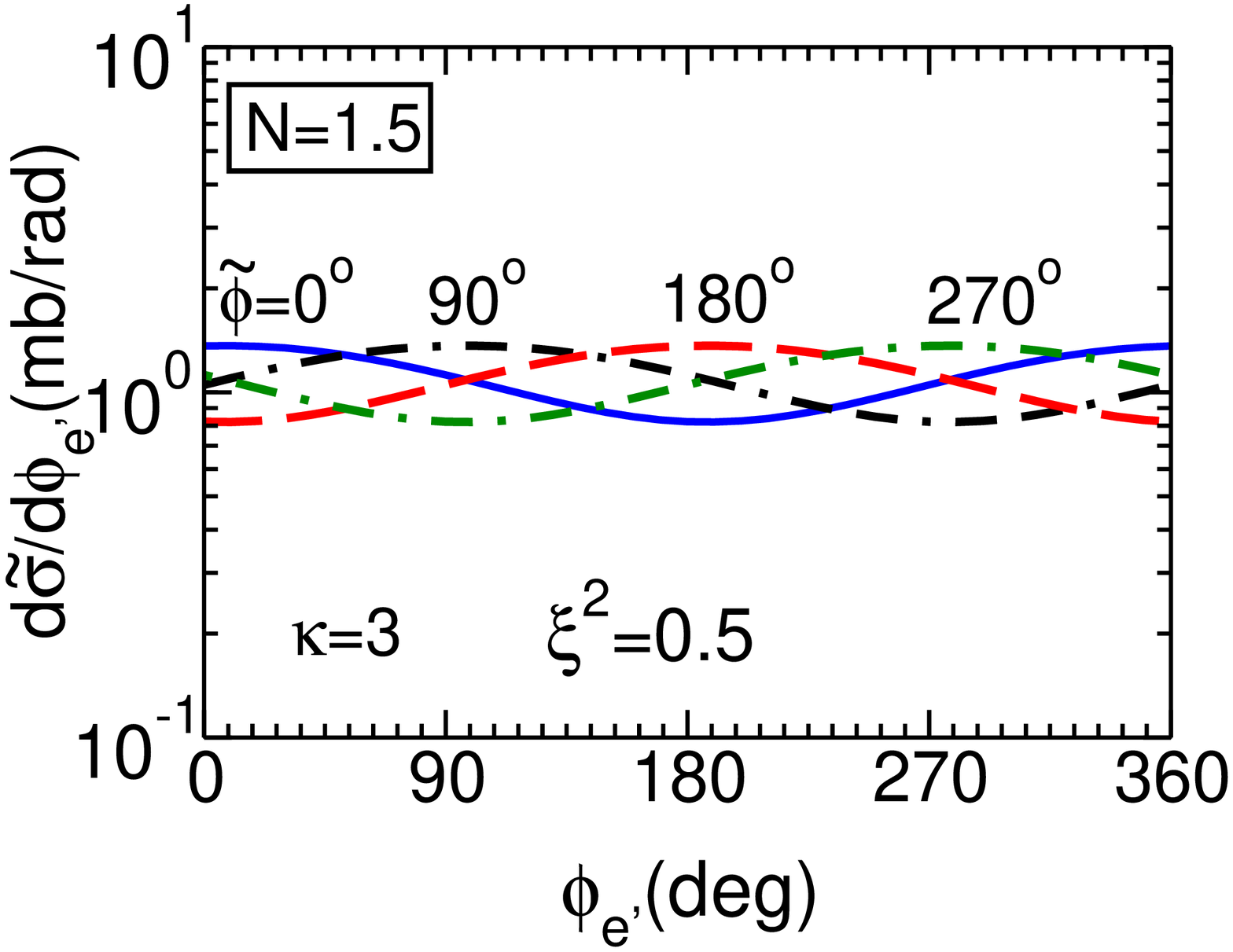}\qquad
\includegraphics[width=0.45\columnwidth]{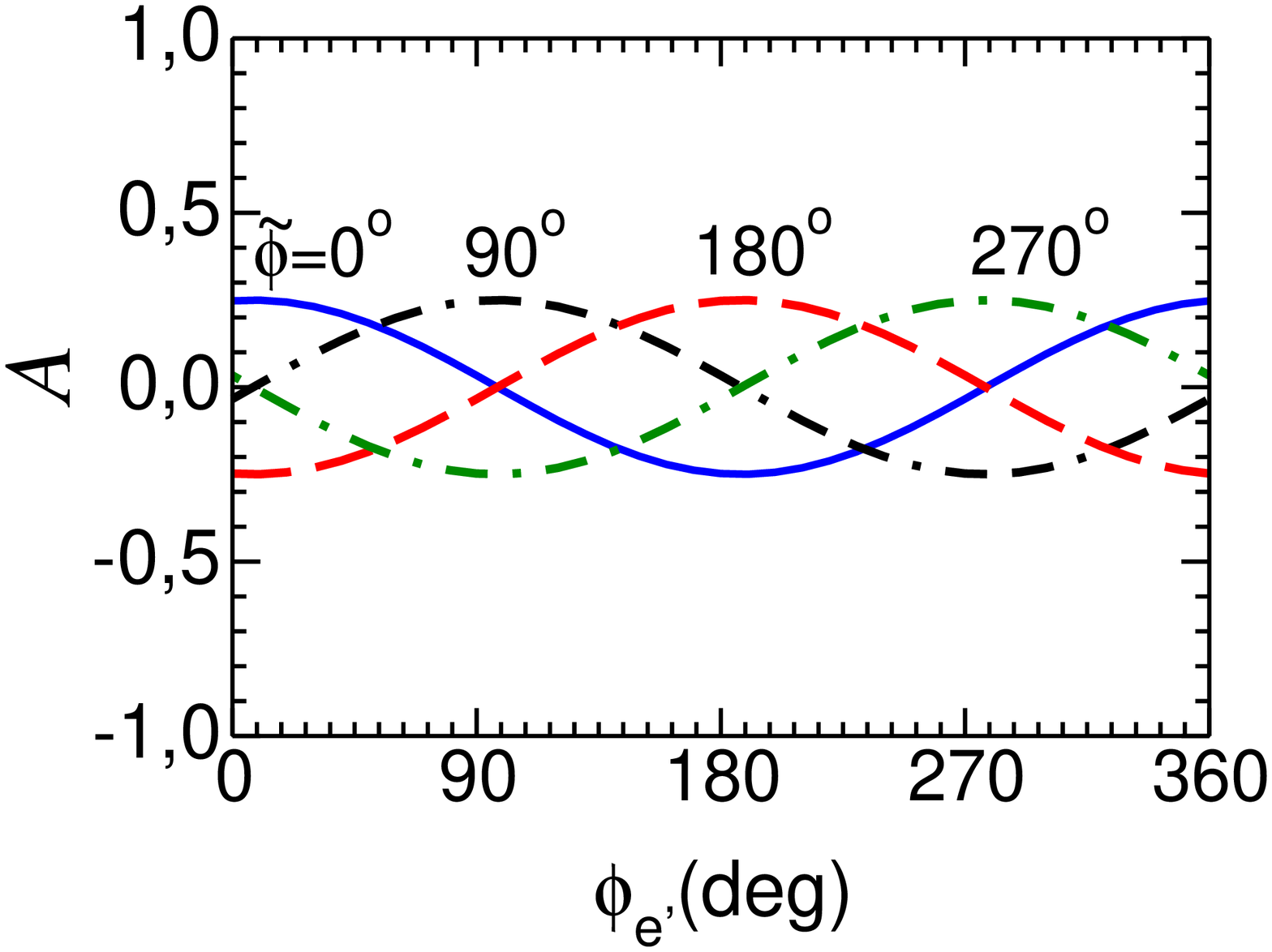}
\includegraphics[width=0.45\columnwidth]{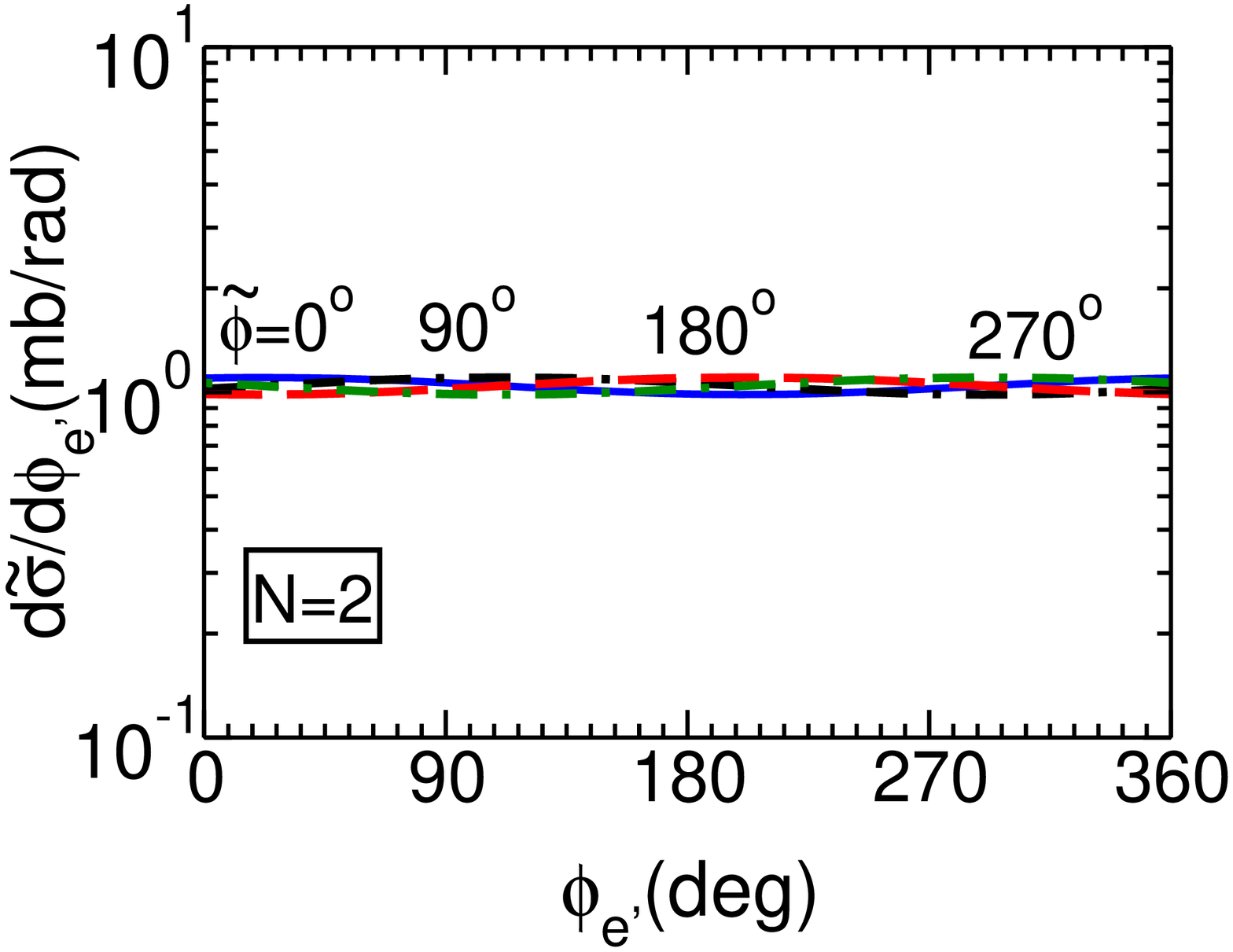}\qquad
\includegraphics[width=0.45\columnwidth]{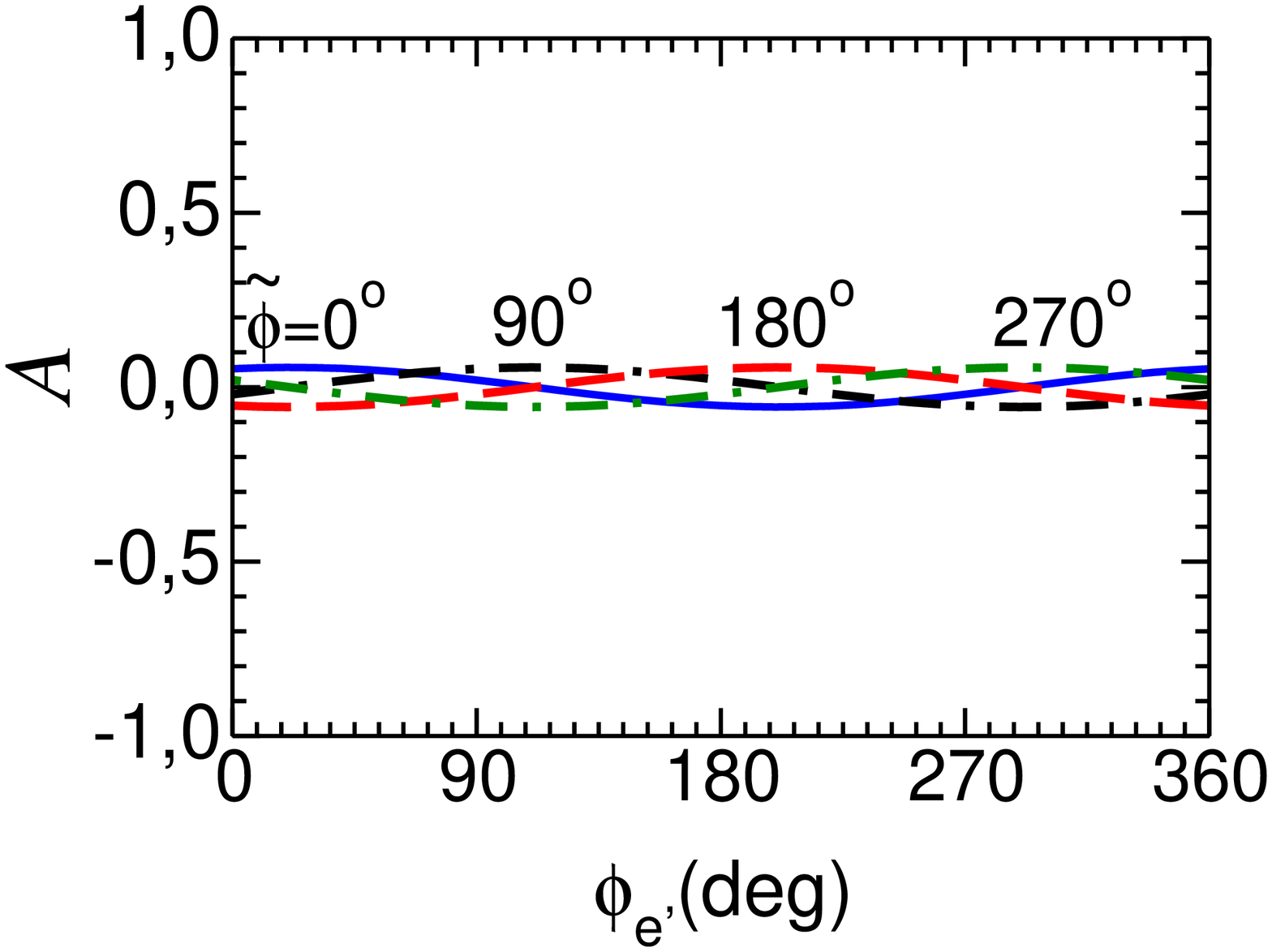}
\caption{\small{
 Left column: The differential cross section (\ref{CPC4}) as
 a function of the azimuthal
 angle of the outgoing electron momentum
 $\phi_{e'}$ for different values of the
 carrier phase $\tilde\phi$ and different numbers of
 cycles in a pulse, $N$ as depicted in the plots.
 The solid, dash-dash-dotted, dashed and dash-dotted
 curves correspond to the carrier phase equal 0,
 90, 180 and 270 degrees, respectively.
 Right column: The anisotropy  as a function
 of $\phi_{e'}$ for different $\tilde\phi$.
 For the hyperbolic secant shape with $N=0.5$;
 $\xi^2=0.5$ and $ \kappa=\omega'/\omega'_1=3$.
 \label{Fig:CEC1}}}
\end{figure}

 The differential cross section (\ref{CPC4})
 as a  function of the azimuthal angle
 $\phi_{e'}$ for different values of the carrier phase $\tilde\phi$
 for the pulses with $N=0.5$, 1, 1.5 and 2 for the hyperbolic secant shape
 with $\kappa=\omega'/\omega'_1=3$ and $\xi^2=0.5$
 is exhibited in the left panels of Fig.~\ref{Fig:CEC1}.

 One can see a clear bump-like structure of the distribution,
 where the bump position coincides with the corresponding value of
 the carrier phase.
 The reason of such behaviour is the same as in case of
 the Breit-Wheeller process, explained by the highly
 oscillating factor in Eq.~(\ref{UU8}) with inequality~(\ref{UU81})
 valid for very short pulses. The impact of CEP decreases with increasing
 pulse duration (or $N$) and becomes very small at $N>2$ which is in agreement
 with prediction of~\cite{CE5} for the linearly polarized pulse.

  Corresponding anisotropies defined  by Eq.~(\ref{U9}) with substitution
  $d\sigma(\phi_e)\to d\sigma(\phi_{e'})$
   are exhibited in the right panels of Fig.~\ref{Fig:CEC1}.
  One can see a strong dependence
  of the anisotropy on the carrier phase for the short pulses which follows to the
  bump-like behavior of the differential cross sections shown in the
  left panels. The maximum effect is expected for the
  sub-cycle pulse with $N=0.5$, where similar to the Breit-Wheeler process, the
  anisotropy takes a maximum value ${\cal A}\simeq 1$
  at $\phi_{e'}=\tilde\phi$ and  $|{\cal A}|<1$ at $\phi_{e'}\neq\tilde\phi$.
  It takes a minimum value ${\cal A}\simeq -1$
  at  $\phi_e-\tilde\phi=\pm\pi$. For the short pulses with $N=2$
  the absolute value of the anisotropy is much lower than 1.

  The CEP effect is sensitive to the sub-threshold parameter
  $\kappa=\omega'/\omega'_1$. Our prediction for
  $\kappa=2$ is exhibited in Fig.~\ref{Fig:CEC2}.

 \begin{figure}[ht]
\includegraphics[width=0.45\columnwidth]{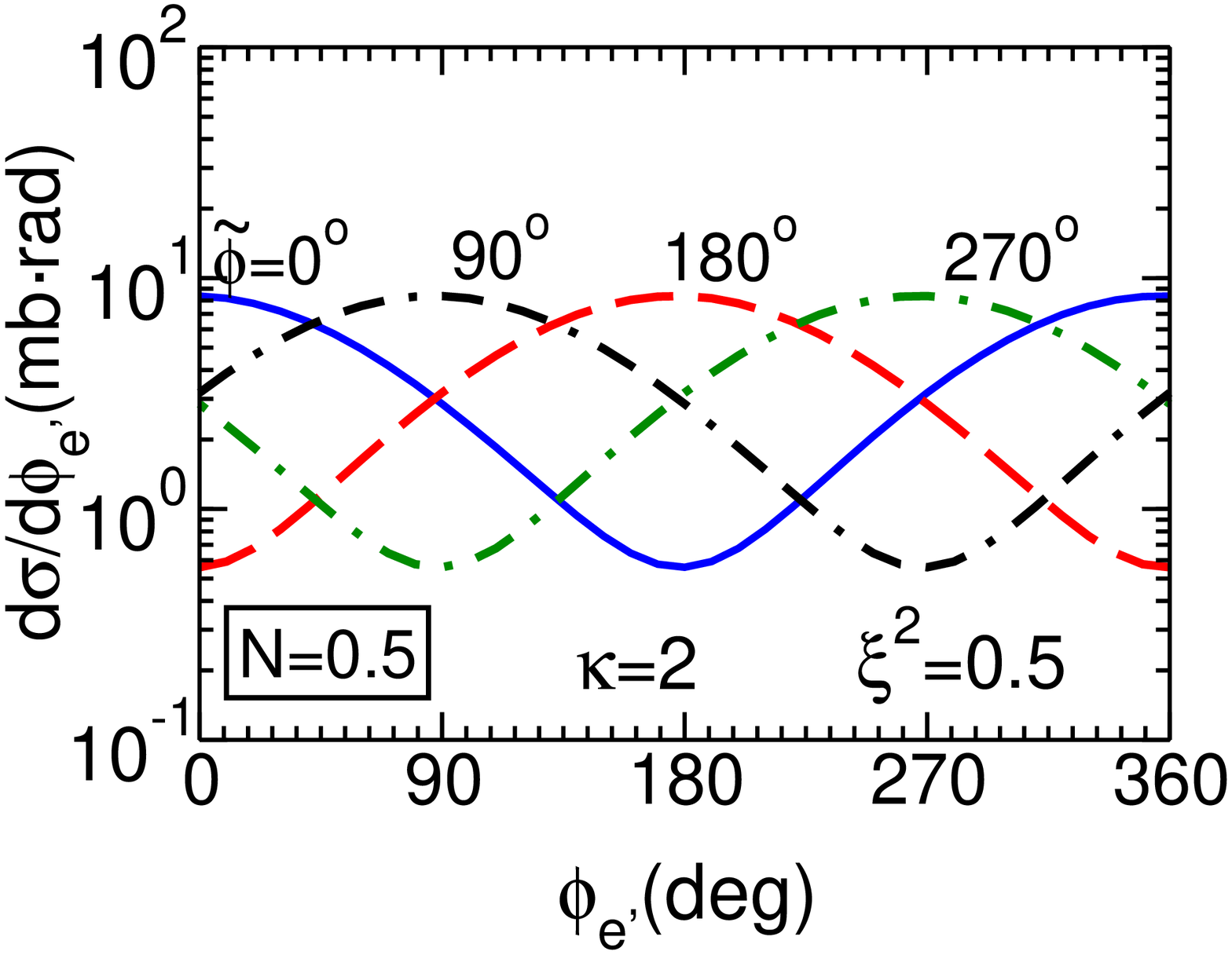}\qquad
\includegraphics[width=0.45\columnwidth]{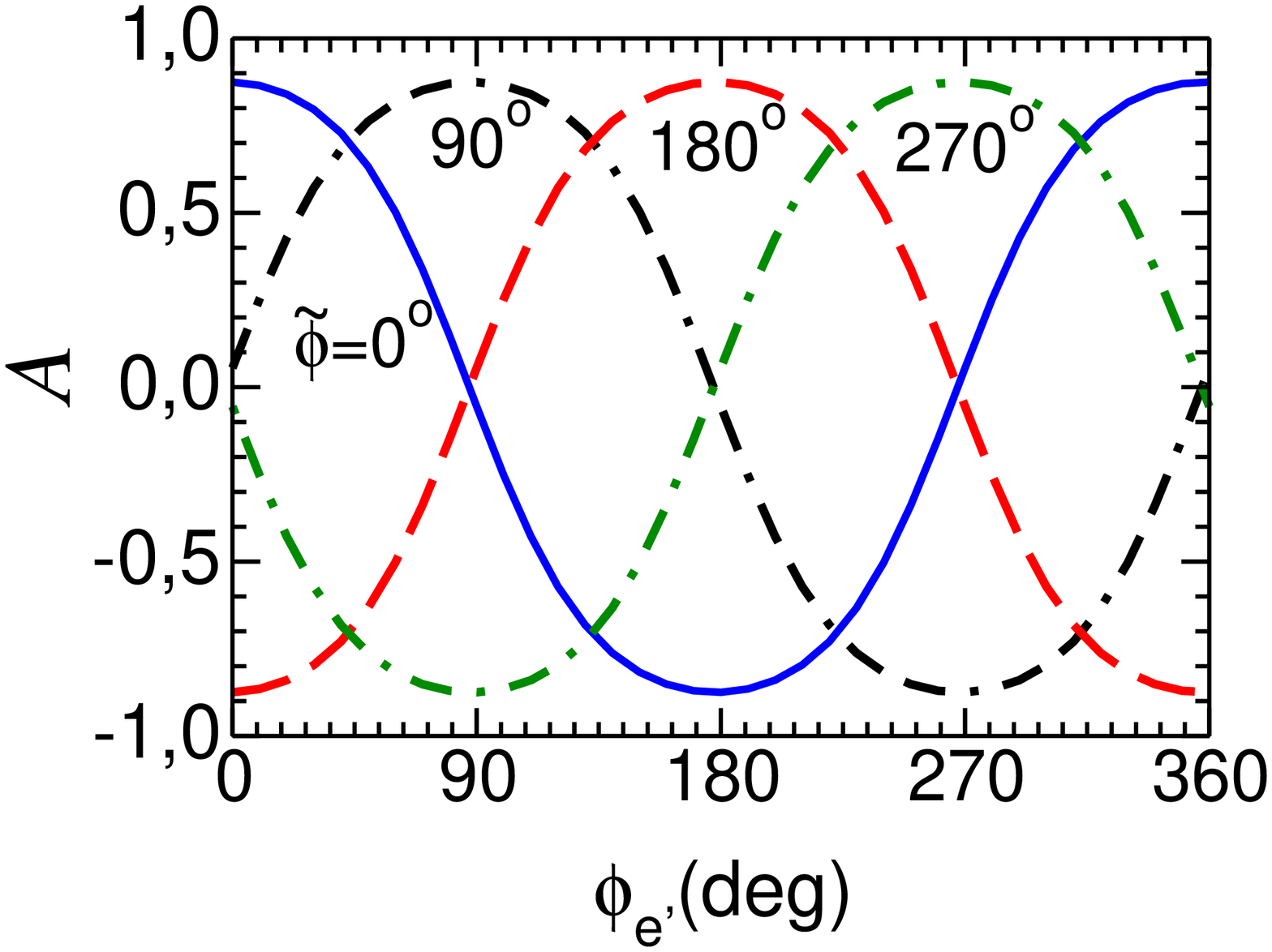}
\includegraphics[width=0.45\columnwidth]{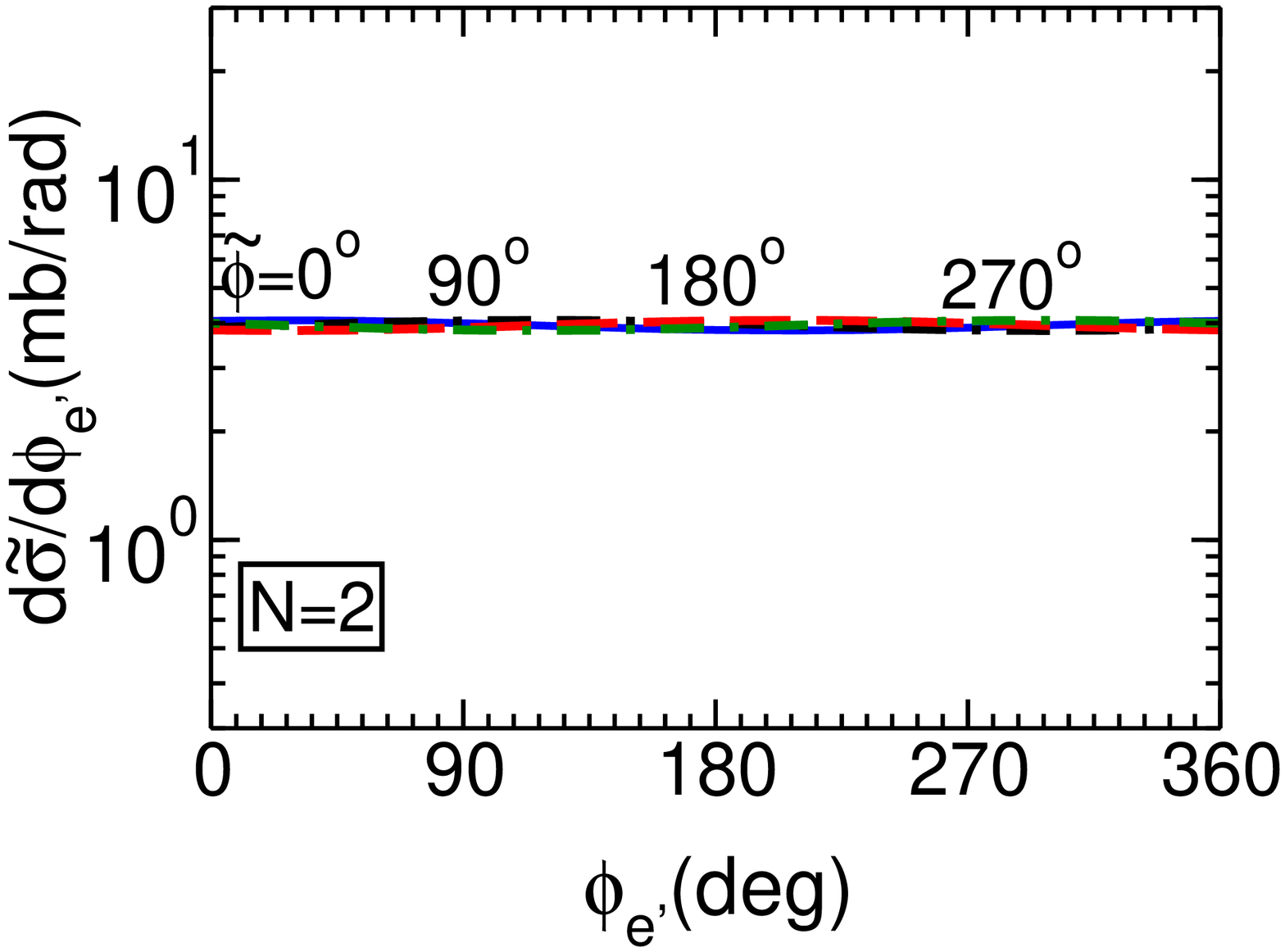}\qquad
\includegraphics[width=0.45\columnwidth]{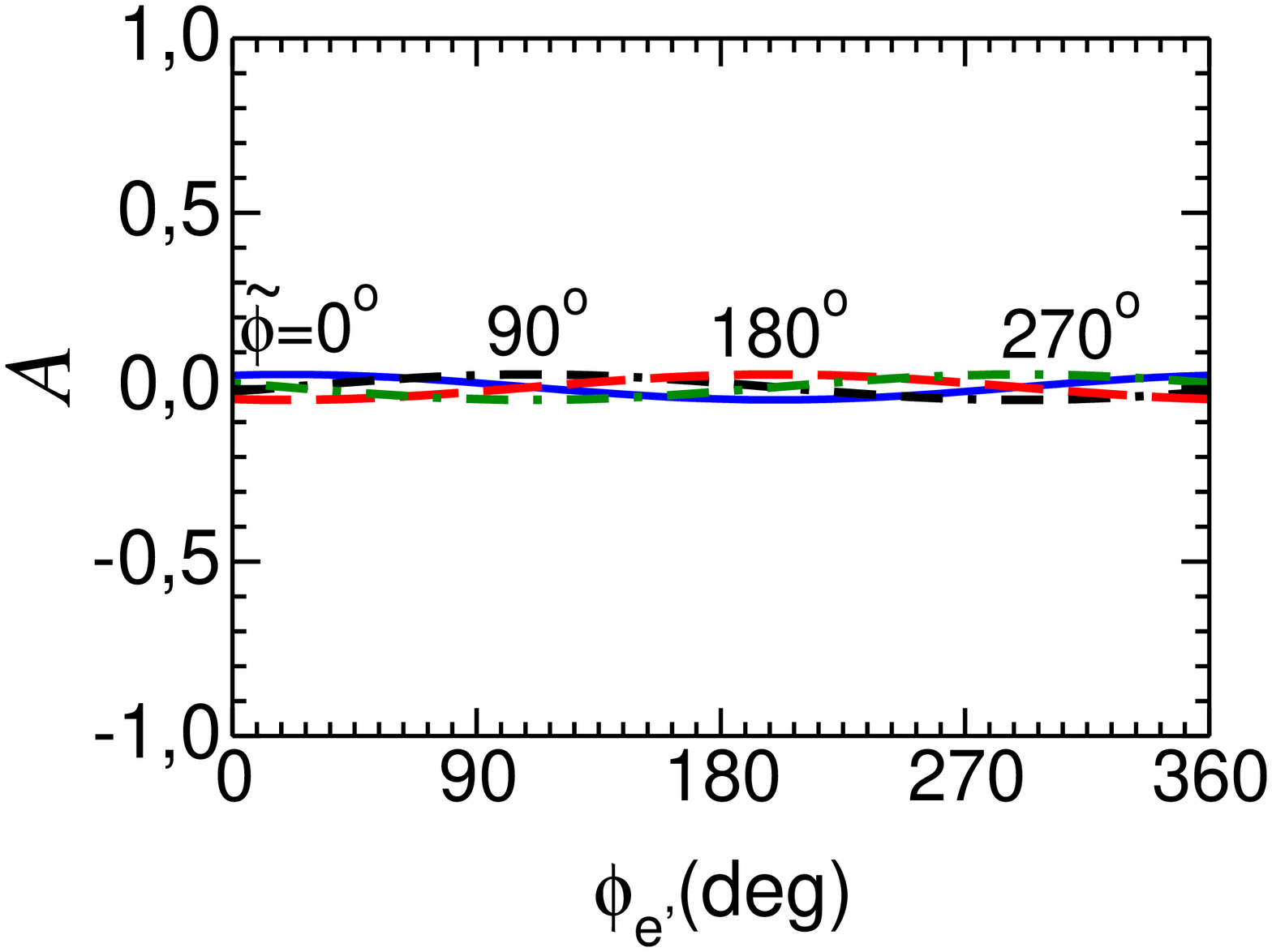}
\caption{\small{The same as in Fig.~\ref{Fig:CEC1} but
 for $\kappa=2$ and for two values of $N=0.5$ (top row)
 and 2 (bottom row).
 \label{Fig:CEC2}}}
\end{figure}

 In this case effect of CEP is smaller.
 Thus, for $N=0.5$ the height of the bumps in the cross sections
 decreases by more than the factor of four compare to the case
 of $\kappa=3$. The impact of CEP practically disappear for $N\geq 2$.

 Finally we, note that similarly to the Breit-Wheeler process,
 the differential cross sections and anisotropies of
 the generalized Compton process are the functions of
 the scale variable $\Phi=\phi_{e'}-\tilde\phi$
 which leads to independence of the corresponding
 observables from $\tilde\phi$ (at fixed $\Phi$),
 and independence of integrated over $\phi_{e'}$
 cross sections from CEP,
 similarly to the results presented in Fig.~\ref{Fig:CE2}.


\section{summary}

 In summary we have considered the carrier envelope phase (CEP) effect
 in a short circularly polarized electromagnetic (laser)
 pulse for  $\ee$ pair production (generalized Breit-Wheeler process)
 and for emission of single photon off an relativistic electron
 (generalized Compton scattering) produced in essentially  multi-photon
 region. In both cases we found a
 strong dependence of the differential cross sections
 as a function of the azimuthal angle of the outgoing particle on CEP.
 In the first case it is the azimuthal angle of the outgoing electron (positron),
 while in the second case it is the azimuthal angle of the recoil electron
 (or photon). For very short pulses
 the corresponding cross sections have a bump-like
 structure where the bump position coincides with the CEP value.
 The height of the bumps
 for sub-cycle pulses with $N\leq1$ reaches orders of magnitude.
 This means that studying the
 azimuthal angle distributions  may be used as a unique
 and power method for the CEP determination in case of
 circularly polarized laser beam. The CEP effect
 decreases quite clearly with increasing the pulse duration and becomes
 negligible for pulses with number of oscillations $N\geq3$.
 Note that this effect becomes relevant in the essentially multi-photon
 (cumulative) region.

\acknowledgments

The authors acknowledge fruitful discussions with
R. Sauerbrey and T. E. Cowan within the HIBEF project.



\begin{thebibliography}{30}
\bibitem{Tajima}
G. A. Mourou, T. Tajima, and S. V. Bulanov.
Rev. Mod. Phys. {\bf 78}, 309 (2006).
%
\bibitem{Piazza}
A.~Di~Piazza, C. M\"uller, K.~Z.~Hatsagortsyan, and
C.~H.~Keitel.
Rev. Mod. Phys. {\bf 84}, 1177 (2012).
%
\bibitem{I-22}
V.~Yanovsky {\it et al.} Opt. Express {\bf 16}, 2109 (2008).

\bibitem{CLF}
 \begin{verbatim} http://www.clf.stfc.ac.uk/CLF/.\end{verbatim}

\bibitem{ELI}
 \begin{verbatim} http://www.eli-beams.eu.\end{verbatim}

\bibitem{hiper}
 \begin{verbatim} http://www.hiper-laser.org. \end{verbatim}

\bibitem{sarov}
\begin{verbatim}
 https://www.ipfran.ru/english/science/las_phys.html.
\end{verbatim}
%
\bibitem{ShortPulse}
 A.~L.~Cavalieri {\it et al.}  New J. Phys. {\bf 9}, 242 (2007);
 Z.~Major {\it et al.}, AIP Conference
Proceedings {\bf 1228}, 117 (2010).

\bibitem{ShortPulse_2}
 { Major~Z} {\it et al.}
 AIP Conference Proceedings {\bf 1228}, 117 (2010)
%
\bibitem{Mackenroth-2011}
  F.~Mackenroth and A.~Di Piazza.
  Phys.\ Rev.\  A {\bf 83}, 032106 (2011).
\bibitem{atto}
X.~Feng, S.~Gilbertson, H.~Mashiko, He~Wang, S.~D. Khan, M.~Chini,
Yi~Wu, K.~Zhao, and Z.~Chang Phys.\ Rev.\ Lett.\ {\bf103}, 183901
(2009).
\bibitem{I-222}
{F.~Krausz, and M.~Ivanov.}
  Rev.\ Mod.\ Phys.\ {\bf 81}, 163 (2009).
%
\bibitem{CE1}
G. G. Paulus {\it et al}. Phys. Rev. Lett. 91, 253004 (2003).

\bibitem{CE2}
T. Wittmann {\it et al}. Nature Phys. 5, 357 (2009).

\bibitem{CE3}
E. Goulielmakis {\it et al}. Science 305, 1267 (2004).

\bibitem{CE4}
 M. Kre\ss{} {\it et al}. Nature Phys. 2, 327 (2006).

\bibitem{CE5}
 F.~Mackenroth, A.~Di~Piazza, and C.~H.~Keitel.
  Phys.\ Rev.\ Lett.\ {\bf 105}, 063903 (2010).
 \bibitem{CE8}
 D.~Seipt and B.~K\"ampfer,
  Phys.\ Rev.\ A {\bf 88}, 012127 (2013).
  \bibitem{CE6}
  K.~Krajewska and J.~Z.~Kaminski.
  Phys.\ Rev.\ A {\bf 86}, 052104 (2012).

\bibitem{CE7}
 F.~Hebenstreit, R. Alkofer, G.~V.~Dunne, and H.~Gies.
  Phys.\ Rev.\ Lett.\ {\bf 102}, 150404 (2009).
\bibitem{A1}
  M.~J.~A.~Jansen and C.~Müller,
  arXiv:1511.07660 [hep-ph].
%
 \bibitem{A2}
 A.~Nuriman, Zi-Liang Li, and Bai-Song Xie,
 Phys. Lett. B, {\bf 726}, 820 (2015).
%

 \bibitem{A4}
 S.~Meuren, C.~H.~Keitel and A.~Di Piazza,
  arXiv:1503.03271 [hep-ph].

 \bibitem{A5}
 S. Meuren, K.~Z. Hatsagortsyan, C.~H. Keitel, and A. Di Piazza,
 Phys. Rev. D {\bf 91}, 013009 (2015).
\bibitem{TitovPRA}
  A.~I.~Titov, B.~K\"ampfer, H.~Takabe and A.~Hosaka.
  Phys.\ Rev.\ A {\bf 87}, 042106 (2013).
 \bibitem{TitovEPJD}
  A.~I.~Titov, B.~K\"ampfer, T.~Shibata, A.~Hosaka and H.~Takabe.
  Eur.\ Phys.\ J.\ D {\bf 68}, 299 (2014).
\bibitem{Ritus-79}
  V.~I.~Ritus. J.~Sov. Laser Res. (United States),
 {\bf 6:5}, 497 (1985).
\bibitem{TitovPRL}
 A.~I.~Titov,  H.~Takabe, B.~K\"ampfer, and A.~Hosaka.
  Phys.\ Rev.\ Lett.~{\bf 108}, 240406 (2012).
 \bibitem{Nousch}
  T.~Nousch, D.~Seipt, B.~K\"ampfer, and A.~I.~Titov.
   Phys.\ Lett.\ B {\bf 715}, 246 (2012).
  %
  \bibitem{Krajewska}
  K.~Krajewska and J.~Z.~Kaminski.
  Phys.\ Rev.\ A {\bf 86}, 052104 (2012).
\bibitem{Boca-2009}
  M.~Boca and V.~Florescu.
  Phys.\ Rev.\  A {\bf 80}, 053403 (2009).
%
\bibitem{Heinzl-2009}
  T.~Heinzl, D.~Seipt, and B.~K\"ampfer.
  Phys.\ Rev.\ A {\bf 81}, 022125 (2010).
\bibitem{Seipt-2011}
  D.~Seipt and B.~K\"ampfer.
  Phys.\ Rev.\  A {\bf 83}, 022101 (2011).
%
\bibitem{Dinu}
  V.~Dinu, T.~Heinzl, and A.~Ilderton.
  Phys.\ Rev.\ D {\bf 86}, 085037 (2012).
%
\bibitem{Seipt-2012}
  D.~Seipt and B.~K\"ampfer.
  Phys.\ Rev.\ D {\bf 85}, 101701 (2012).
  %
  \bibitem{Krajewska-2012}
  K.~Krajewska and J.~Z.~Kaminski.
  Phys.\ Rev.\ A {\bf 85}, 062102 (2012).
%
 \bibitem{A3}
 K. Krajewska, F. Cajiao Vélez, and J. Z. Kamiński,
 Phys. Rev. A {\bf 91}, 062106 (2015).
 \bibitem{LL4}
  V.~B.~Berestetskii, E.~M.~Lifshitz, and L.~P.~Pitaevskii. {\it
 Quantum Electrodynamics}. 2nd ed., (Course of theoretical
 physics; vol.~4), Oxford, New York, Pergamon Press Ltd.\ (1982).
%
\bibitem{DSeipt-2014}
D. Seipt and B. K\"ampfer. Phys. Rev. A, {\bf 89}, 023433, (2014).
 \end{thebibliography}
\end{document}